\documentclass[%
 reprint,
amsmath,amssymb,
prab,
nofootinbib,
]{revtex4-2}

\usepackage{dcolumn}
\usepackage{bm}
\usepackage{siunitx}
\usepackage{graphicx}
\usepackage{dcolumn}
\usepackage{bm}

\usepackage{comment}

\begin{document}

\preprint{APS/123-QED}

\title{Intrabeam scattering studies with large-emittance-ratio ion beams in the Relativistic Heavy Ion Collider, and implications for the Electron-Ion Collider}

\author{Y. Luo}
\email{yluo@bnl.gov}
\author{B. Lepore}%
\author{K. Mernick}%
\author{W. Bergan}%
\author{S. Nagaitsev}%
\affiliation{%
 Brookhaven National Laboratory, Upton NY, 11973, USA
}%

\date{\today}

\begin{abstract}

The Electron–Ion Collider (EIC), to be constructed at Brookhaven National Laboratory, will collide polarized, high-energy electron beams with hadron beams, achieving peak luminosities of up to $1.0 \times 10^{34}$ cm$^{-2}$s$^{-1}$. To reach such high luminosity, the EIC will employ flat-beam collisions at the interaction point. The design transverse emittance ratio will be about 10:1 in the Hadron Storage Ring (HSR).  Thanks to stochastic cooling and precise decoupling, /we successfully generated and accelerated gold ion beams with a large emittance ratio of 11:1 in the Relativistic Heavy Ion Collider (RHIC). In this article, we present results of intrabeam scattering (IBS)  measurements and modeling for large-emittance gold ion beams, both without and with controlled betatron coupling. To model the IBS growth, we use the formulas developed by Lebedev and Nagaitsev.

\end{abstract}



\maketitle


\section{Introduction}

The Electron–Ion Collider (EIC), to be constructed at Brookhaven National Laboratory (BNL), will collide polarized, high-energy electron beams with hadron beams, achieving peak luminosities of up to $1.0 \times 10^{34}$ cm$^{-2}$s$^{-1}$ over a center-of-mass energy range of 20–140 GeV~\cite{EIC}. The EIC consists of two storage rings: the Hadron Storage Ring (HSR) and the Electron Storage Ring (ESR), both of which will be housed in the existing tunnel of the Relativistic Heavy Ion Collider (RHIC)~\cite{RHIC}. The HSR will reuse the arcs of the RHIC yellow ring and connect them with new straight sections. RHIC was decommissioned in February 2026.

To reach high luminosities, the EIC will employ flat beams at the interaction point (IP). Flat ion beams can be realized through unequal transverse emittances and $\beta$ functions at the IP, with $\epsilon_x / \epsilon_y$ and $\beta_x^* / \beta_y^*$ both on the order of 10~\cite{EIC}. Owing to stochastic cooling and precise decoupling, we demonstrated a transverse emittance ratio of 11:1 in RHIC in 2023 using gold ion beams~\cite{PRL_paper, PRAB_paper1}. In 2025, we further demonstrated acceleration of large-emittance-ratio ion beams from 31 to 100 GeV/nucleon~\cite{PRAB_paper2}. These results validate the HSR design assumption that large-emittance-ratio hadron beams can be generated at injection and preserved during acceleration to collision energy.

The availability of flat beams in RHIC enables a range of EIC-relevant beam physics studies. In this work, we present intrabeam scattering (IBS) measurements and modeling for gold ion beams with large transverse emittance ratios, both without and with controlled betatron coupling. To our knowledge, this is the first experimental study benchmarking IBS growth for hadron beams with large emittance ratios, as well as the first to include deliberately introduced betatron coupling.

IBS-induced emittance growth has been extensively studied since Piwinski~\cite{IBS1}. The most widely used formulas, due to Piwinski and Bjorken–Mtingwa~\cite{IBS2}, are derived from two-body Coulomb scattering. More recently, Lebedev and Nagaitsev developed an alternative set of formulas based on the Landau collision integral from plasma physics~\cite{IBS3}. Combined with the coupling parameterization by Lebedev and Bogacz~\cite{coupling0}, this approach enables IBS calculations including betatron coupling. In the uncoupled limit, these formulas reduce to the Bjorken–Mtingwa expressions, but are written in terms of the elliptic integral of the second kind, which can be efficiently evaluated using the method outlined in Ref.~\cite{IBS7}.

Although Piwinski also derived IBS growth-rate expressions including betatron coupling~\cite{IBS5}, their practical use is limited by the complexity of the eigenvectors and the double integrals involved. For electron storage rings with synchrotron radiation, Kubo and Oide proposed a beam-envelope approach~\cite{IBS6} that accounts for betatron coupling, although it does not explicitly provide the IBS diffusion matrix.

For high-energy hadrons (e.g., 100 GeV/nucleon in RHIC), IBS growth is usually dominated by the longitudinal and horizontal planes, with horizontal growth driven by dispersion along the ring. In contrast, direct vertical IBS growth is negligible due to the small vertical dispersion; instead, vertical emittance growth arises primarily through betatron coupling~\cite{IonRun12}.

Consequently, many previous IBS studies for hadron beams assumed either round beams or equal transverse growth rates (“fully coupled” conditions)~\cite{IBS-exp0, IBS-exp1, IBS-exp2, IBS-exp3}. For beams with large emittance ratios, however, such assumptions are no longer valid. In our experiments, the horizontal emittance exceeds the vertical by approximately an order of magnitude. To generate and maintain such beams, residual coupling must be carefully minimized.

Under these conditions, horizontal and vertical IBS growth must be treated independently. Experimentally, we observe distinct growth rates in the two transverse planes, which also depend on the ion beam energy. Accordingly, we abandon the fully coupled approximation. In this work, we employ the Lebedev–Nagaitsev formulas to treat the growth of transverse eigenmodes separately and consistently, both without and with controlled betatron coupling.

The paper is organized as follows. We first review betatron coupling theory, the Lebedev–Nagaitsev IBS formulas, and the experimental setup. We then present IBS measurements and modeling for well-decoupled beams with large transverse emittance ratios at 31 and 100 GeV/nucleon. Next, we analyze IBS growth of ion beams with large emittance ratios under controlled weak coupling at 31 GeV/nucleon. Finally, we estimate IBS growth times in the EIC hadron storage ring (HSR) at various beam  energies  with different flatness and weak betatron coupling.


\section{Theories and Experiment Setup}

\subsection{Betatron Coupling}

There are two main factors affecting the preservation of a large transverse emittance ratio in the HSR/EIC: betatron coupling and IBS. Betatron coupling resonance couples the horizontal and vertical motions of particles, leading to changes in the betatron tunes and the mixing of transverse emittances~\cite{Coupling1, Coupling2}. The sources of betatron coupling include detector solenoids, skew quadrupoles, and residual coupling sources such as quadrupole roll errors and vertical closed orbits in sextupoles. 

For random sources of betatron coupling in a ring, Hamiltonian perturbation theory is more suitable for analytical estimates than the strict matrix-based approach. In the following, we use the amplitude of the coupling coefficient defined in Hamiltonian perturbation theory to measure the strength of global coupling, which is defined as ~\cite{Coupling2}
\begin{equation}  \label{eq:1}
\begin{aligned}
  C^{-} &= \frac{1}{2\pi} \oint{ \sqrt{\beta_{x} \beta_{y}} \left[ k_{1s}  +  k_s \left(\frac{\alpha_x}{\beta_x} -\frac{\alpha_y}{\beta_y}\right) \right.}\\
  & {\left. - i k_s\left(\frac{1}{\beta_x}+\frac{1}{\beta_y}\right) \right] e^{i (\Psi_{x}-\Psi_{y})}dl}.
\end{aligned}
\end{equation}
Here $k_{1s}$ and $k_s$ are the strengths of skew quadrupoles and solenoids, respectively, and $\beta_{x,y}$ and $\alpha_{x,y}$ are the unperturbed Twiss parameters, $\Psi_{x,y}$ are the unperturbed betatron pahase advances. 

With betatron coupling, both horizontal and vertical motions are composed of two transverse eigenmodes, with eigentunes $Q_{1,2}$ and eigen emittances $\epsilon_{1,2}$. For weak coupling, eigenmode 1 is typically associated with the horizontal plane, and eigenmode 2 with the vertical plane.

The eigentunes can be analytically evaluated through perturbation theory
\begin{equation}  \label{eq:2}
  Q_{1} =  Q_{x,0} - \frac{\Delta}{2} +\frac{1}{2}\sqrt{\Delta^{2}+|C^{-}|^{2}},
\end{equation}
\begin{equation}  \label{eq:3}
 Q_{2} = Q_{y,0} + \frac{\Delta}{2} -\frac{1}{2}\sqrt{\Delta^{2}+|C^{-}|^{2}} .
\end{equation}
Here $Q_{x,y,0}$ are the fractional uncoupled tunes, and $\Delta= Q_{x,0}-Q_{y,0}$. The fractional eigentune split with coupling is 
\begin{equation}  \label{eq:4}
|Q_1 -Q_2| = \sqrt{ \Delta^2 + |C^{-}|^2 }.
\end{equation}
The minimum tune split is equal to the amplitude of the coupling coefficient $|C^{-}|$ when $\Delta=0$. $\Delta=0$ corresponds to the difference coupling resonance. In some literature, $|C^{-}|$ is referred to as the minimum tune split $|\Delta Q_{\min}|$.

Betatron coupling also modifies transverse beam sizes along the ring. Knowing the eigen-emittances $\epsilon_{1,2}$, the beam size matrix with coupling can be calculated as~\cite{Coupling3, Coupling4}
\begin{equation}  \label{eq:5}
  \Sigma_{\mathbf{X}} = \mathbf{P}
         \left( 
         \begin{array}{cccc}
           \epsilon_{1,\mathrm{rms}}  &  0  & 0   & 0 \\
           0  &    \epsilon_{1,\mathrm{rms}}   & 0  & 0  \\
           0  &  0 &    \epsilon_{2,\mathrm{rms}}   & 0  \\
           0  &  0 & 0  &   \epsilon_{2,\mathrm{rms}}  \\
         \end{array}
         \right)     \mathbf{P}^{t}, 
\end{equation}
where $\mathbf{P}$ contains Edwards--Teng's Twiss and coupling parameters:
\begin{equation}   \label{eq:6}
\mathbf{P}=
\left(
\begin{array}{cccc} 
r \sqrt{\beta_1}              &  0  &  c_{11} \sqrt{\beta_2}- \frac{c_{12} \alpha_2}{\sqrt{\beta_2}}  &   \frac{c_{12}}{\sqrt{\beta_2}}   \\
-\frac{\alpha_1 r}{\sqrt{\beta_1}}   &  \frac{r}{\sqrt{\beta_1}}  &  c_{21} \sqrt{\beta_2} - \frac{c_{22} \alpha_2}{\sqrt{\beta_2}}  & \frac{c_{22}}{\sqrt{\beta_2}} \\
 -\frac{c_{12} \alpha_1}{\sqrt{\beta_1}} -c_{22} \sqrt{\beta_1}  & \frac{c_{12}}{\sqrt{\beta_1}} & r \sqrt{\beta_2} & 0 \\
\frac{c_{11} \alpha_1}{\sqrt{\beta_1}} +c_{21} \sqrt{\beta_1} & -\frac{c_{11}}{\sqrt{\beta_1}} & - \frac{\alpha_2 r}{\sqrt{\beta_2}} & \frac{r}{\sqrt{\beta_2}}
\end{array}
\right) .
\end{equation}
In the  betatron parameterization by Lebedev and Bogacz~\cite{coupling0}, $\mathbf{P}$ is denoted as $\mathbf{V}$, since it can be directly constructed from the eigenvectors of the one-turn map, too.  With betatron coupling, the eigen-emittances cannot be directly measured. In the following experiments, knowing the eigen-emittances from modeling, we can use Eq.~(\ref{eq:6}) to calculate transverse beam sizes under coupled situation.

\subsection{Lebedev--Nagaitsev Formulas}

Both Piwinski and Bjorken--Mtingwa derived IBS formulas starting from first principles of two-body Coulomb collisions. Recently, Lebedev and Nagaitsev developed another set of IBS formulas based on the Landau collision integral from plasma physics~\cite{IBS3}. For a uniform, nonrelativistic plasma with a Gaussian distribution, the time derivatives of RMS velocities in the beam frame are
\begin{eqnarray}   \label{eq:7}
            \frac{d}{dt} \left( \begin{array}{c}
                    \sigma^2_{vx} \\
                    \sigma^2_{vy} \\
                    \sigma^2_{vz} \\
                    \end{array}
              \right)  
             & =      & \frac{ (2 \pi)^{3/2} e^4 n L_c }{m^2 \sqrt{\sigma^2_{vx} + \sigma^2_{vy} +  \sigma^2_{vz} }} \nonumber  \\  
             &        & \left( \begin{array}{c}
                                                                   \Psi(\sigma_{vx}, \sigma_{vy} , \sigma_{vz} )\\
                                                                    \Psi(\sigma_{vy}, \sigma_{vz} , \sigma_{vx} )\\
                                                                    \Psi(\sigma_{vz}, \sigma_{vx} , \sigma_{vy} ) \\
                               \end{array}
                       \right).
\end{eqnarray}
with
\begin{equation}  \label{eq:8}
  \begin{array}{lll}
    \Psi(x,y,z)& =&  \frac{\sqrt{2} r}{3 \pi} ( y^2 R_D(z^2,x^2,y^2) + z^2 R_D(x^2,y^2,z^2)  \\
    &  & -2x^2 R_D(y^2,z^2,x^2)  ),
    \end{array}
\end{equation}
\begin{equation}  \label{eq:81}
     r = \sqrt{x^2+y^2+z^2}.
\end{equation}
Here $n$ is the plasma density, $m$ and $e$ are the charged particle's mass and charge, and $L_c$ is the Coulomb logarithm. $R_D (x,y,z)$ is the elliptic integral of the second kind.

To apply the above equations to IBS growth in accelerators, we express the RMS velocities in the beam frame in terms of the RMS momenta ($p_x$, $p_y$, $p_z$) in the laboratory frame, from which the derivatives of the eigen-emittances can  be derived. Using the 4-D betatron coupling parameterization by Lebedev and Bogacz, Lebedev and Nagaitsev obtained 
\begin{equation}  \label{eq:9}
  \begin{array}{lll}
    \frac{d \epsilon_q}{dt} &=& \frac{N r^2_0 c}{3 \sqrt{\pi} \beta^3 \gamma^4 C \epsilon_1 \epsilon_2 \sigma_p } [ L_c \sigma_1 \sigma_2 \sigma_3 (    \\
    &  & ~~\sigma_1^2 R_D(\sigma^2_2,\sigma^2_3, \sigma^2_1)  ( Trace(\mathbf{a}^{q})-3[\mathbf{T}^t\mathbf{a}^{q}\mathbf{T}]_{11} )   \\
    &  & + \sigma_2^2  R_D(\sigma^2_3,\sigma^2_1, \sigma^2_2)  ( Trace(\mathbf{a}^{q})-3[\mathbf{T}^t\mathbf{a}^{q}\mathbf{T}]_{22} ) \\
    &  & + \sigma_3^2  R_D(\sigma^2_1,\sigma^2_2, \sigma^2_3)  ( Trace(\mathbf{a}^{q})-3[\mathbf{T}^t\mathbf{a}^{q}\mathbf{T}]_{33} ) \\
    &    & )  ] , \\ 
  \end{array}
\end{equation}
\begin{equation}   \label{eq:10}
  \begin{array}{lll}
    \frac{d \sigma^2_p}{dt} &=& \frac{N r^2_0 c}{3 \sqrt{\pi} \beta^3 \gamma^4 C \epsilon_1 \epsilon_2 \sigma_p } [ L_c \sigma_1 \sigma_2 \sigma_3 (    \\
    &  & ~~\sigma_1^2 R_D(\sigma^2_2,\sigma^2_3, \sigma^2_1)  ( Trace(\mathbf{a}^{s})-3[\mathbf{T}^t\mathbf{a}^{s}\mathbf{T}]_{11} )   \\
    &  & + \sigma_2^2  R_D(\sigma^2_3,\sigma^2_1, \sigma^2_2)   ( Trace(\mathbf{a}^{s})-3[\mathbf{T}^t\mathbf{a}^{s}\mathbf{T}]_{22} ) \\
      &  & + \sigma_3^2 R_D(\sigma^2_1,\sigma^2_2, \sigma^2_3)   ( Trace(\mathbf{a}^{s})-3[\mathbf{T}^t\mathbf{a}^{s}\mathbf{T}]_{33} ) \\
    &    & )  ] . \\
  \end{array}
\end{equation}
Eqs.~(\ref{eq:9}) and (\ref{eq:10}) are reproduced here from Eqs.~(53) and (54) in Ref.~\cite{IBS3}. Here $N$ is the particle population per bunch, $r_0$ is the classical radius of the charged particle, and $c$ is the speed of light. $\beta$ and $\gamma$ are relativistic parameters. $\epsilon_{q}$, with $q=1,2$, are the transverse eigen-emittances, and $\sigma_p$ is the RMS momentum spread. $\sigma_{1,2,3}$  are uncoupled RMS momenta in the laboratory frame.  For definitions of $\mathbf{T}$, $\mathbf{a}^{q}$, and $\mathbf{a}^{s}$, see Ref.~\cite{IBS3}. Note that the matrix $\mathbf{a}^{q}$ has the dimension of \textit{length}, while both $\mathbf{T}$ and $\mathbf{a}^{s}$ are dimensionless.  These quantities are expressed in terms of coupled optics parameters, which connect eigen-emittances to RMS values of momenta. 

To obtain Eqs.~(\ref{eq:9}) and (\ref{eq:10}), a Gaussian distribution in all 6 dimensions is assumed. For weakly coupled beams, the distribution remains close to Gaussian, which justifies the use of these formulas in the present study. As mentioned earlier, in the absence of coupling, the Lebedev--Nagaitsev IBS formulas reduce to the Bjorken--Mtingwa's, but expressed in terms of $R_D$ \cite{IBS7}, which accelerates the evaluation of IBS growth rates.

\subsection{Experiment Setup}

In RHIC, the transverse emittances were measured using ionization profile monitors (IPMs), one for the horizontal plane and one for the vertical plane. They were installed at different locations in the ring, with horizontal and vertical $\beta$-functions of approximately 200~m, respectively. The IPMs directly measure the transverse beam profiles and fit them to Gaussian distributions to derive the RMS beam sizes. Together with  the model $\beta$-functions at the IPM locations, the transverse emittances are reported. The main uncertainty in the IPM emittance measurement arises from the uncertainty in the $\beta$-functions at the IPMs. There is approximately a 10\% $\beta$-beat uncertainty from the lattice model in RHIC. In this article, 95\% normalized transverse emittances are used for experiment data, which is 
\begin{equation}
  \epsilon_{95\%, norm}  = 6 \epsilon_{norm} = 6 \beta \gamma \epsilon_{rms}.
  \end{equation}
Here the RMS emittance $\epsilon_{rms,x,y} =  \sigma_{x,y}^2 / \beta_{x,y} $, with  $\sigma_{x,y}$ and $\beta_{x,y}$ as transverse beam size and $\beta$ function.

The bunch length was measured using a wall current monitor (WCM) in RHIC. In the following experiments, we used 28 equally spaced gold ion ($^{197}$Au$^{79+}$) bunches to improve the quality of the IPM emittance measurements. The longitudinal profile of each bunch can be directly measured. In RHIC, the bunch length is typically reported using the full width at half maximum (FWHM). In the following experiments, a Gaussian distribution is assumed, for which the FWHM bunch length is 2.355 times the RMS bunch length. According to Ref.~\cite{IBS-exp1}, the uncertainty in the RHIC bunch length measurement is about 0.25~ns, or 7.5~cm. The typical RMS ion bunch length in these experiments is greater than 1.4~m.

In 2006, we successfully demonstrated and implemented a global decoupling feedback system at RHIC~\cite{CouplingFB1,CouplingFB2}. This system relies on continuous measurement of the coupling coefficient, achieved through a baseband phase-locked loop tune meter. The coupling coefficient is derived from the measured eigenmode amplitude projections onto the transverse axes. Since the coupling coefficient is a complex number, we sorted the existing skew quadrupoles at RHIC into two orthogonal families to effectively correct the global coupling. Since then, decoupling feedback has become an integral part of routine RHIC operation, applied during injection, at the store, and during acceleration. In RHIC, after applying decoupling feedback, the amplitude of the coupling coefficient $|C^{-}|$ is typically about 0.001.

To generate large emittance ratios for IBS studies, stochastic cooling was used in our following experiments. Operational longitudinal cooling of bunched ion beams was first achieved in RHIC in 2007~\cite{StochasticCooling1}, followed by vertical cooling in 2011. In 2012, with the addition of the horizontal plane, full three-dimensional (3-D) stochastic cooling became operational~\cite{StochasticCooling2}. To maximize the transverse emittance ratio, horizontal-plane cooling is turned off while vertical-plane cooling is maintained. In this configuration, IBS increases the horizontal emittance while stochastic cooling suppresses the vertical emittance, thereby maximizing the transverse emittance ratio~\cite{PRAB_paper1}. 

\section{IBS for Well-decoupled Beams} 

\begin{figure}[bht]
  \centering
\includegraphics*[width=60mm,angle=-90]{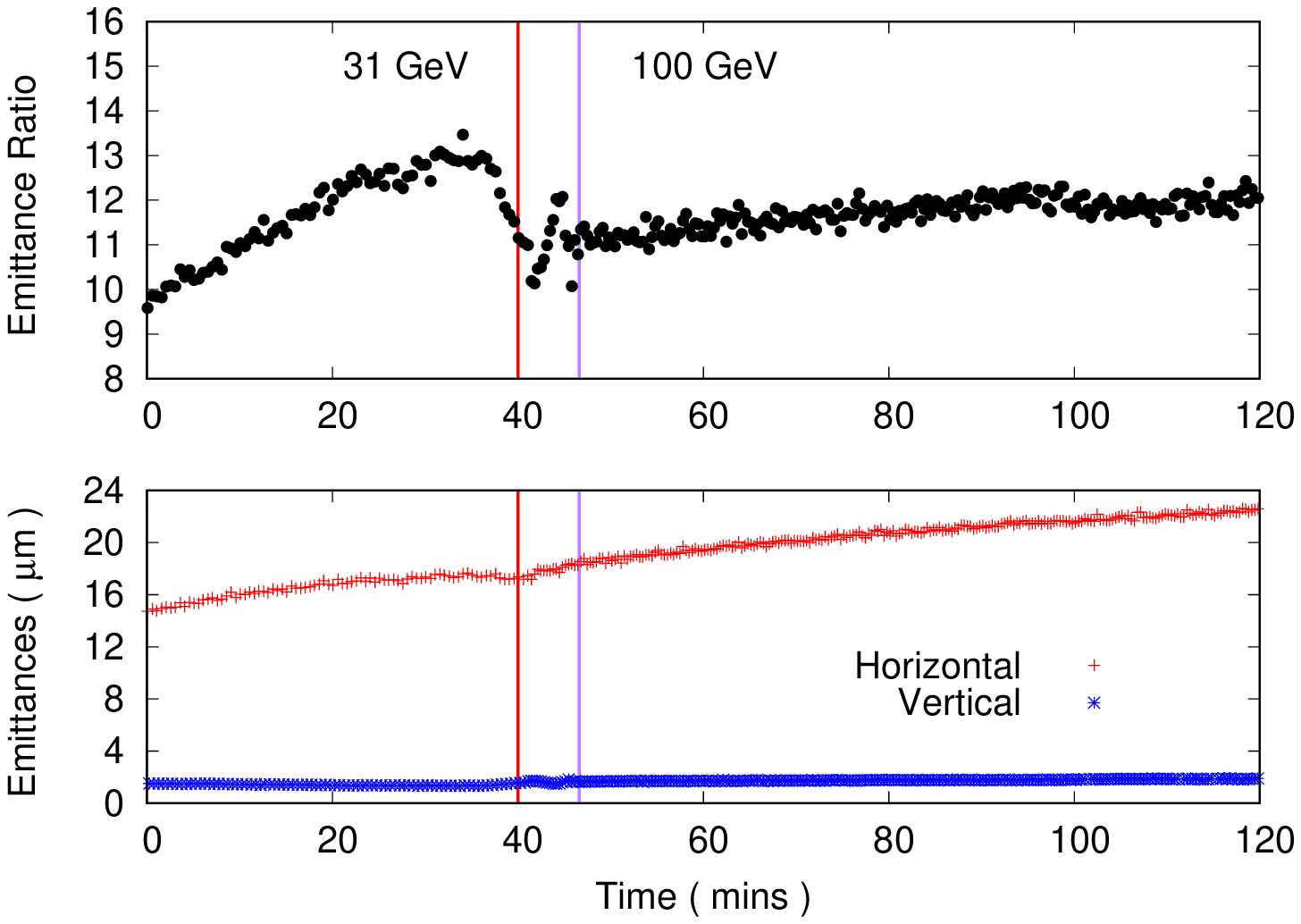}
\caption{\label{fig:1} Measured  transverse  emittances in the bottom plot and the transverse emittance ratio in the top plot for the experiment  conducted on July 23, 2025.}
\end{figure}

For the well-decoupled case, we use the measurement data from a recent beam experiment of accelerating large-emittance ion beams in RHIC, conducted on July 23, 2025. Figure~\ref{fig:1} shows the measured transverse emittances $\epsilon_{x,y}$ and the emittance ratio $r = \epsilon_x / \epsilon_y$ during the entire process. Between the two vertical lines, the beam was accelerated from 31~GeV/nucleon to 100~GeV/nucleon.

In this experiment, 28 gold ion bunches were used in the Yellow ring of RHIC. The average ion population was about $0.6\times 10^9$ per bunch prior to acceleration. At 31~GeV/nucleon, we first generated a large emittance ratio by applying only vertical stochastic cooling to reduce the vertical emittance while allowing the horizontal emittance to grow due to IBS. In the following experiments, it typically takes about 2.5 hours to reach a transverse emittance ratio of approximately 10:1.

When the emittance ratio reached about 11:1, the beam was accelerated to 100~GeV/nucleon. With high-performance control of orbit, tune, and coupling during the energy ramp, the large emittance ratio was successfully preserved. At 100~GeV/nucleon, the large-emittance-ratio beam was maintained for approximately 2 hours.

The transverse fractional tunes were kept at (0.238, 0.212) at 31~GeV/nucleon and during the acceleration to 100~GeV/nucleon. At store, these tunes were maintained for the first 50 minutes. The tunes were then moved to the HSR design values (0.238, 0.210), where the beam was kept for approximately another hour.

\begin{figure}[bht]
  \centering
\includegraphics*[width=60mm,angle=-90]{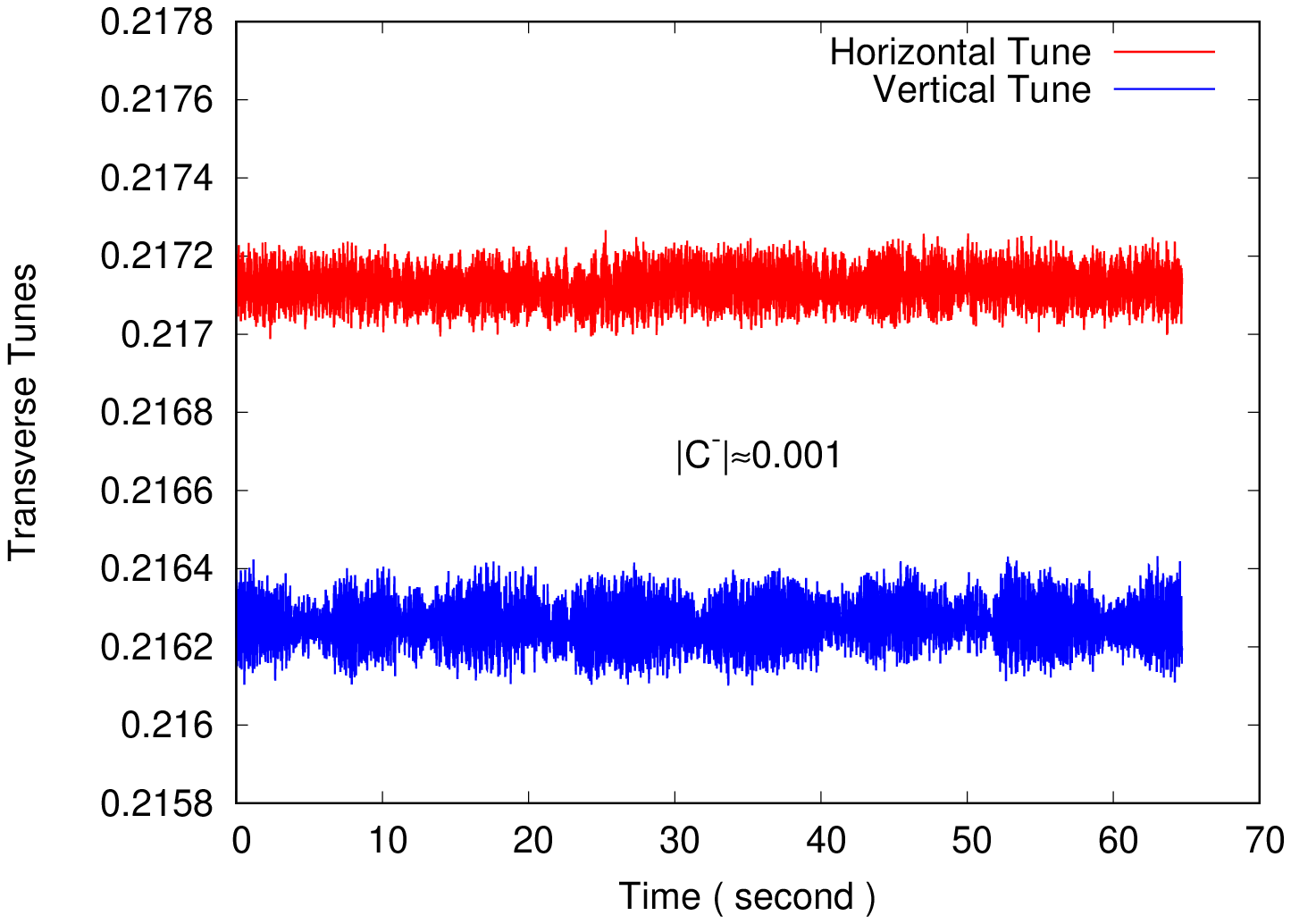}
\caption{\label{fig:2} Betatron tunes during  measurement of the minimum tune split, or the amplitude of coupling coefficient $|C^{-}|$,  at 100 GeV /nucleon.}
\end{figure}

Throughout the experiment, betatron coupling was corrected using decoupling feedback.  In this experiment, we measured the minimum tune split, i.e., the amplitude of the coupling coefficient $|C^{-}|$, at 100~GeV/nucleon before the end of this experiment, as shown in Fig.~\ref{fig:2}. Since the tune split was 0.026 during cooling, acceleration, and at the beginning of store, which  was more than an order of magnitude larger than the residual coupling amplitude, we assumed that the Yellow ring was well decoupled during this experiment.

\subsection{Modeling At 31 GeV / nucleon}

Just before accelerating the large-emittance-ratio beam, we intentionally paused for about 5 minutes without any cooling. The initial purpose was to ensure that stochastic cooling was completely switched off. Here, we model the evolution of the transverse emittances and bunch length due to IBS during this period. Table~\ref{tab:1} lists the initial measured beam parameters used for IBS modeling, along with the averaged IBS growth times obtained from both measurement and modeling.

For IBS modeling, the RMS momentum spread is required but cannot be measured directly. In this experiment, 400~kV of 28~MHz RF cavities was used. Given the measured RMS bunch length and the lattice model, the RMS momentum spread can be determined via single-particle tracking around the ring with the RF cavities on.

\begin{figure*}
  \centering
  \includegraphics*[width=100mm,angle=-90]{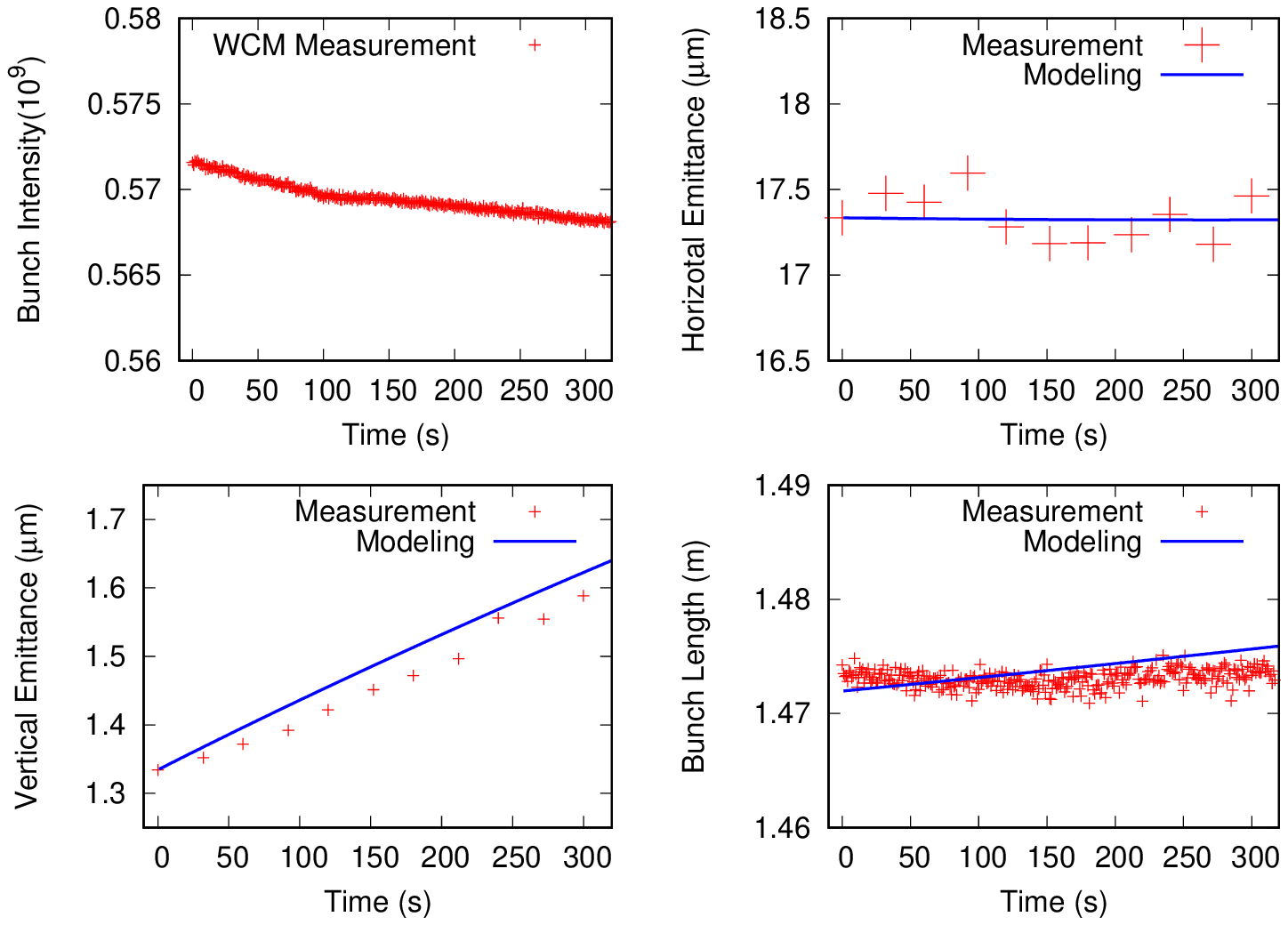}
\caption{\label{fig:3} Emittances and bunch length from measurement and IBS molding at 31 GeV/nucleon. Top-left: averaged  bunch intensity. Top-right: horizontal emittance. Bottom-left: vertical emittance. Bottom-Right: RMS bunch length. }
\end{figure*}

During these 5 minutes at 31~GeV/nucleon, we observed about a 19.1\% increase in the vertical emittance, while the horizontal emittance and bunch length remained nearly unchanged. The transverse emittance ratio decreased from 13.5:1 to about 11:1. Figure~\ref{fig:3} shows the averaged bunch intensity, transverse emittances, and bunch length. The measurement data are shown in red, and the IBS modeling results are shown in blue.

In these experiments, we do not attempt to compare instantaneous IBS growth times between measurement and modeling, as such quantities cannot be determined precisely and typically carry large uncertainties. Instead, the evolution of the emittances and bunch length due to IBS, from both measurement and modeling, can be best fitted with
\begin{equation}
y = A + B \exp(t / \tau),
\end{equation}
where $A$, $B$, and $\tau$ are fitting parameters. We focus on the net changes over a finite observation period. For simplicity, we define an average growth time $\overline{\tau}$ using the initial and final values of a quantity as
\begin{equation}
y_{1} = y_{0} \exp(\Delta t / \overline{\tau}),
\end{equation}
where $y_{0}$ and $y_{1}$ are the initial and final values, and $\Delta t$ is the duration of the observation period. From the measurements, the average growth time for the vertical emittance is 25.1~min. The growth times for the horizontal emittance and bunch length are $-10$~h and $+239$~h, respectively, and can be considered effectively infinite.

As shown in Fig.~\ref{fig:3}, there is negligible bunch intensity loss during these 5 minutes, and it is therefore neglected in the IBS modeling. In the simulation, the IBS growth rates, emittances, bunch length, and momentum spread are updated every second. The modeling reproduces the measurements well. From the simulation, the average growth time for the vertical emittance is about 25.7~min, in good agreement with the measured value. The growth times for the horizontal emittance and bunch length are effectively negative and positive infinity, respectively. Table~\ref{tab:1} summarizes the averaged IBS growth times from both measurement and modeling.

\begin{figure*}
  \centering
  \includegraphics*[width=100mm,angle=-90]{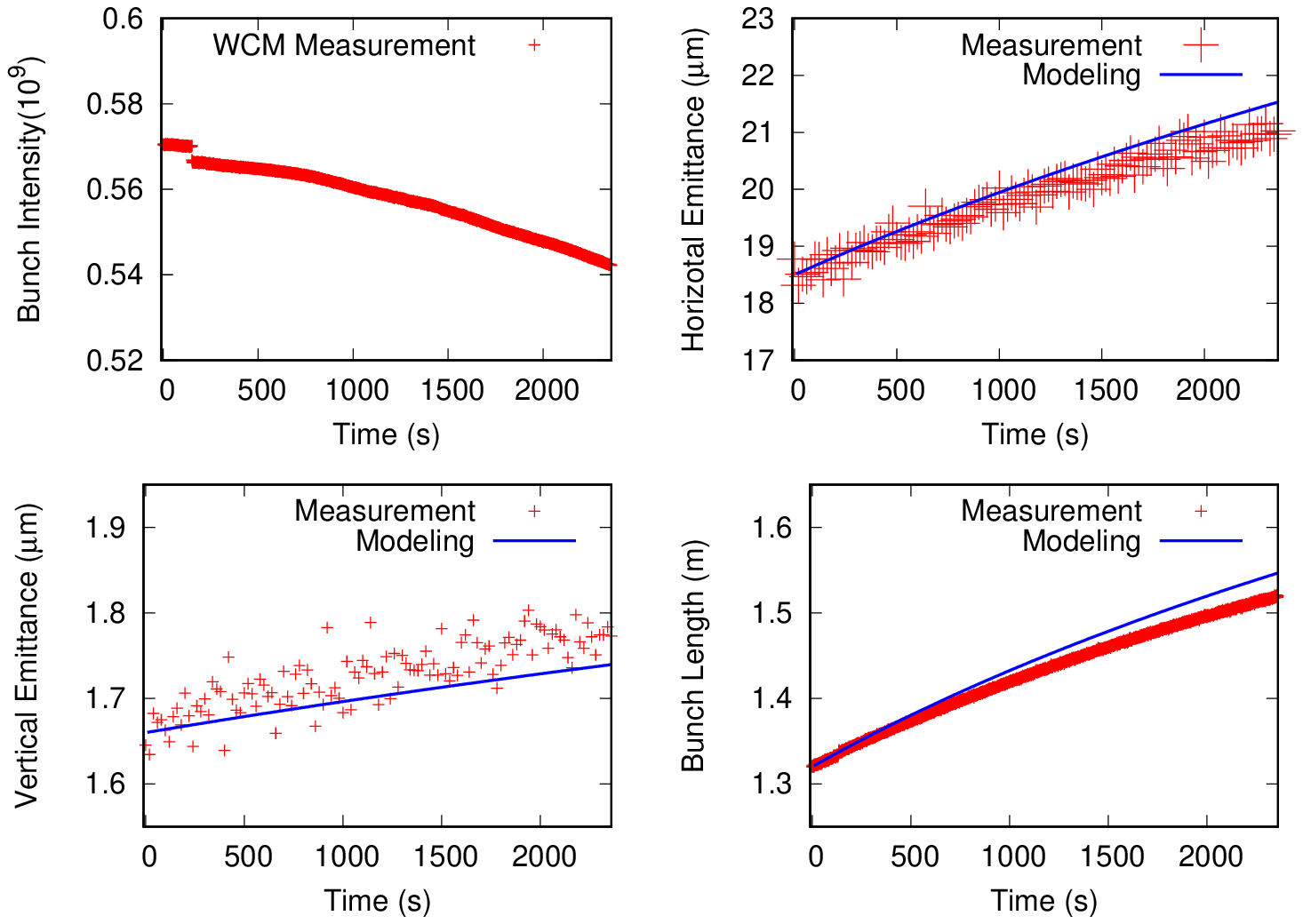}
\caption{\label{fig:4}  Emittances and bunch length from measurement and IBS molding at 100 GeV/nucleon. Top-left: averaged  bunch intensity. Top-right: horizontal emittance. Bottom-left: vertical emittance. Bottom-Right: RMS bunch length. }
\end{figure*}

\begin{table} 
	\vspace*{-.8\baselineskip}
	\centering 
	\caption{\label{tab:1} Initial beam parameters  for IBS modeling and IBS  growth times from both measurement and modeling at 31 GeV/nucleon}
	\begin{tabular}{lcc} 
		\hline \hline
		\bf quantity                &    \bf unit          &  \bf  value   \\ 
		\hline 
		\multicolumn{3}{l}{\bf Initial parameters:} \\
                Bunch intensity        &       $10^{9}$     &      0.57     \\
                Horizontal emittance   &     \textmu m     &     17.3      \\
                Vertical emittance     &     \textmu m     &     1.33      \\
		Bunch length           &        m            &      1.47  \\
		Momentum spread        &       $10^{-3}$     &       1.30 \\ \hline
                \multicolumn{3}{l}{\bf IBS growth times from measurement :}\\
                Horizontal emittance           &   hour              &      -10.0         \\
                Vertical emittance                       &   minute              &     25.1            \\
                Bunch length                   &   hour                  &   239         \\  \hline
                \multicolumn{3}{l}{\bf IBS growth times from modeling:}\\
                Horizontal emittance                    &   hour              &       -167         \\
                Vertical emittance                      &   minute            &      25.7          \\
                Bunch length                  &   hour              &        31.6           \\ \hline \hline
	\end{tabular} 
\end{table}

\subsection{Modeling At 100 GeV / nucleon}

Next, we model the IBS measurement data at 100~GeV/nucleon. Small discontinuities are observed in the measured transverse emittances and bunch length during the tune change from (0.238, 0.212) to (0.228, 0.210). For consistency, we use only the data from the first 40 minutes, during which the tunes were kept at (0.238, 0.212).

Figure~\ref{fig:4} shows the measured bunch intensity, transverse emittances, and bunch length as red dots, together with the IBS modeling results in blue. Unlike the 31~GeV/nucleon case, clear growth is observed in both transverse emittances and the bunch length. The averaged growth times for the horizontal and vertical emittances are 4.9~h and 11.1~h, respectively, indicating that the horizontal emittance grows roughly twice as fast as the vertical. The averaged growth time for the bunch length is 4.8~h, comparable to that of the horizontal emittance.

During this 40-minute period, the bunch intensity decreased by about 6\%, as measured by the wall current monitor (WCM), while the beam current measured by the direct current current transformer (DCCT) remained nearly constant. This indicates that the beam loss was primarily due to ions leaking out of the longitudinal RF bucket. For IBS modeling, the WCM bunch intensity is fitted with a piecewise exponential function, before and after 1440~s. In the simulation, the IBS growth rates, emittances, bunch length, and momentum spread are updated every 10~s.

Table~\ref{tab:2} lists the initial beam parameters used for IBS modeling and the averaged growth times from both measurement and simulation. As shown in Fig.~\ref{fig:4}, the IBS model reproduces the measured evolution of the emittances and bunch length reasonably well. The modeled averaged growth times for the horizontal and vertical emittances are 4.3~h and 13.2~h, respectively, while the bunch length growth time is 4.2~h, somewhat shorter than the measured 4.8~h. This discrepancy may be attributed to ions leaking out of the RF bucket, which effectively slows the measured bunch length growth relative to the model prediction.

\begin{table} 
	\vspace*{-.8\baselineskip}
	\centering 
	\caption{\label{tab:2} Initial beam parameters  for IBS modeling and IBS  growth times from both measurement and modeling at 100 GeV/nucleon}
	\begin{tabular}{lcc} 
		\hline \hline
		\bf quantity                &    \bf unit          &  \bf  value   \\ 
		\hline 
		\multicolumn{3}{l}{\bf initial parameters:} \\
                Bunch intensity        &       $10^{9}$     &      0.57     \\
                Horizontal emittance   &       \textmu m     &    18.50 \\
                Vertical emittance     &       \textmu m     &    1.66 \\
		Bunch length           &        m            &      1.32  \\
		Momentum spread        &       $10^{-4}$     &       5.35 \\ \hline
                \multicolumn{3}{l}{\bf IBS time from measurement :}\\
                Horizontal emittance                     &   hour              &    4.9            \\
                Vertical emittance                       &   hour              &    11.1         \\
                Bunch length                             &   hour              &    4.8          \\  \hline
                \multicolumn{3}{l}{\bf IBS time from modeling:}\\
                Horizontal emittance                     &   hour              &    4.3         \\
                Vertical emittance                       &   hour              &    13.2         \\
                Bunch length                             &   hour              &    4.2     \\ \hline \hline
	\end{tabular} 
\end{table}

\section{IBS for Weakly coupled Beams}

In this section, we study the growth of transverse emittances and bunch length due to IBS under weakly coupled conditions. For this purpose, a dedicated beam experiment was performed on Nov.~14, 2025. To avoid ion leakage from the RF buckets observed at 100~GeV/nucleon, the experiment was conducted at 31~GeV/nucleon, where the bunch length growth is negligible, as discussed above.

\begin{figure}[bht]
  \centering
\includegraphics*[width=60mm,angle=-90]{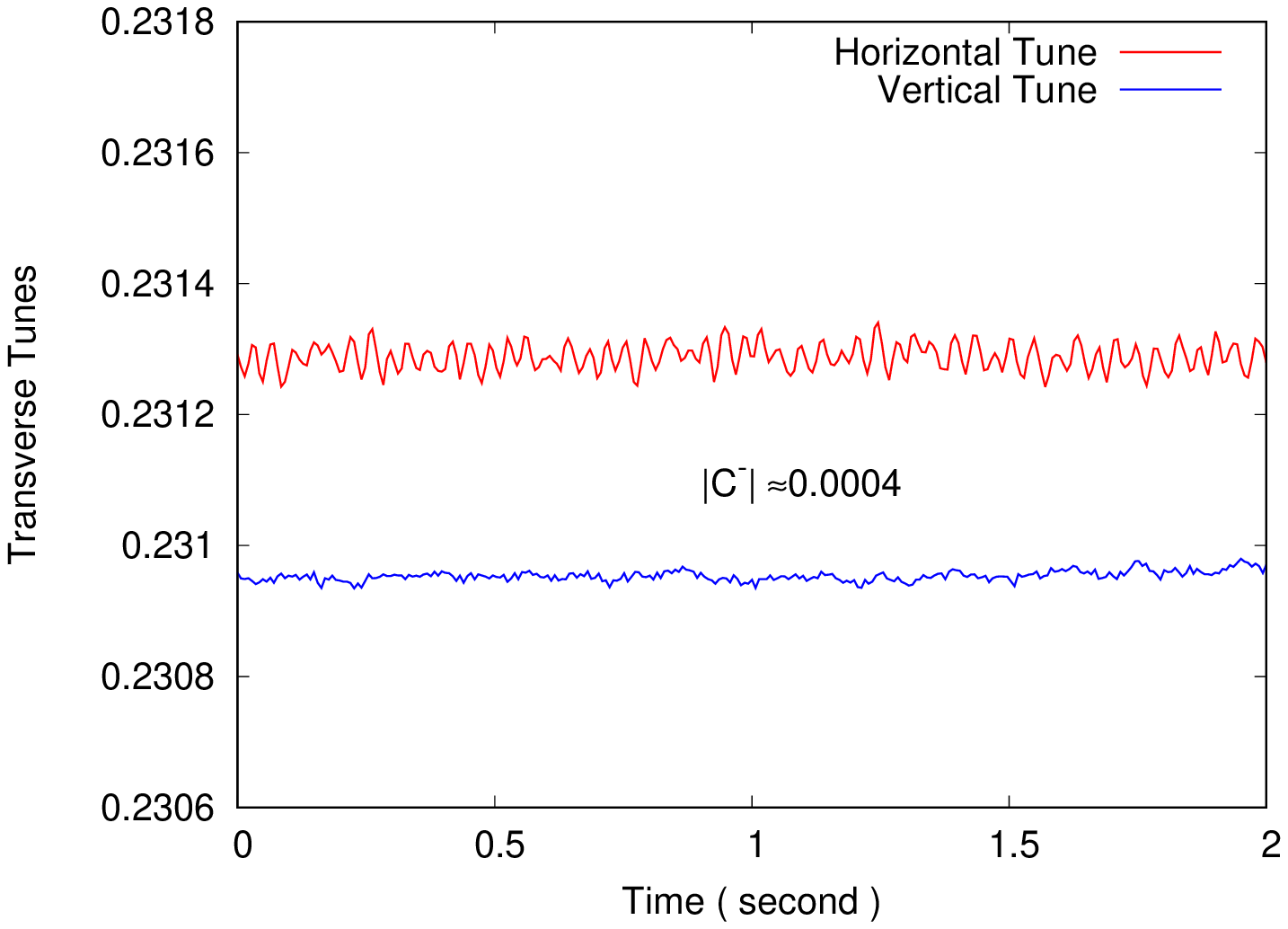}
\caption{\label{fig:40} Betatron tunes during  measurement of the minimum tune split, or the amplitude of coupling coefficient $|C^{-}|$  at 31 GeV /nucleon during the experiment conducted on Nov. 14, 2025.}
\end{figure}

In this experiment, 28 gold-ion bunches were stored in the Yellow ring at an average bunch intensity of $0.7 \times 10^{9}$. Prior to cooling, the minimum tune split—corresponding to the amplitude of the betatron coupling coefficient $|C^{-}|$—was measured to verify effective decoupling. The betatron tunes were brought as close as possible by adjusting the arc quadrupole strengths. As shown in Fig.~\ref{fig:40}, a minimum tune split of approximately $4 \times 10^{-4}$ was achieved, indicating a well-decoupled lattice. During the subsequent IBS experiment, the uncoupled betatron tunes were set to (0.235, 0.215).

Measurements of emittances and bunch length due to IBS were performed under three conditions: well decoupled, $|C^{-}| = 0.01$, and $|C^{-}| = 0.02$. For each case, the vertical emittance was first cooled to obtain an emittance ratio of approximately 10:1 under well-decoupled conditions. Cooling was then turned off, and coupling was introduced by ramping the skew-quadrupole currents. The beam evolution was recorded for approximately 10 minutes. After each measurement, the skew quadrupoles were reset, and the beam was recooled to prepare for the next condition.

Each RHIC ring contains three families of skew quadrupoles (F1, F2, and F3), each powered by four independent supplies driving a total of 16 magnets. In this experiment, the F1 and F3 families were used to introduce controlled coupling on top of a well-decoupled lattice.

In the absence of coupling, the ionization profile monitors (IPMs) directly measure the horizontal and vertical (eigen-)emittances. Under coupled conditions, however, the IPMs measure RMS beam sizes, and the emittances are inferred using the uncoupled lattice functions $\beta_{x,y,0}$. These inferred quantities do not correspond to eigen-emittances and are therefore referred to here as quasi-emittances. The corresponding beam sizes at the IPMs, including betatron coupling, can be reconstructed from the quasi-emittances and the uncoupled lattice functions.

Although the eigen-emittances cannot be directly measured during the coupled periods, their changes due to IBS and coupling can be inferred from the difference between the eigen-emittances measured immediately before coupling is introduced and after it is removed. Here, we assume that the eigen-emittances remain adiabatic invariants during the ramping of the skew-quadrupole currents.

\begin{figure} [hbt]
  \centering
\includegraphics*[width=60mm,angle=-90]{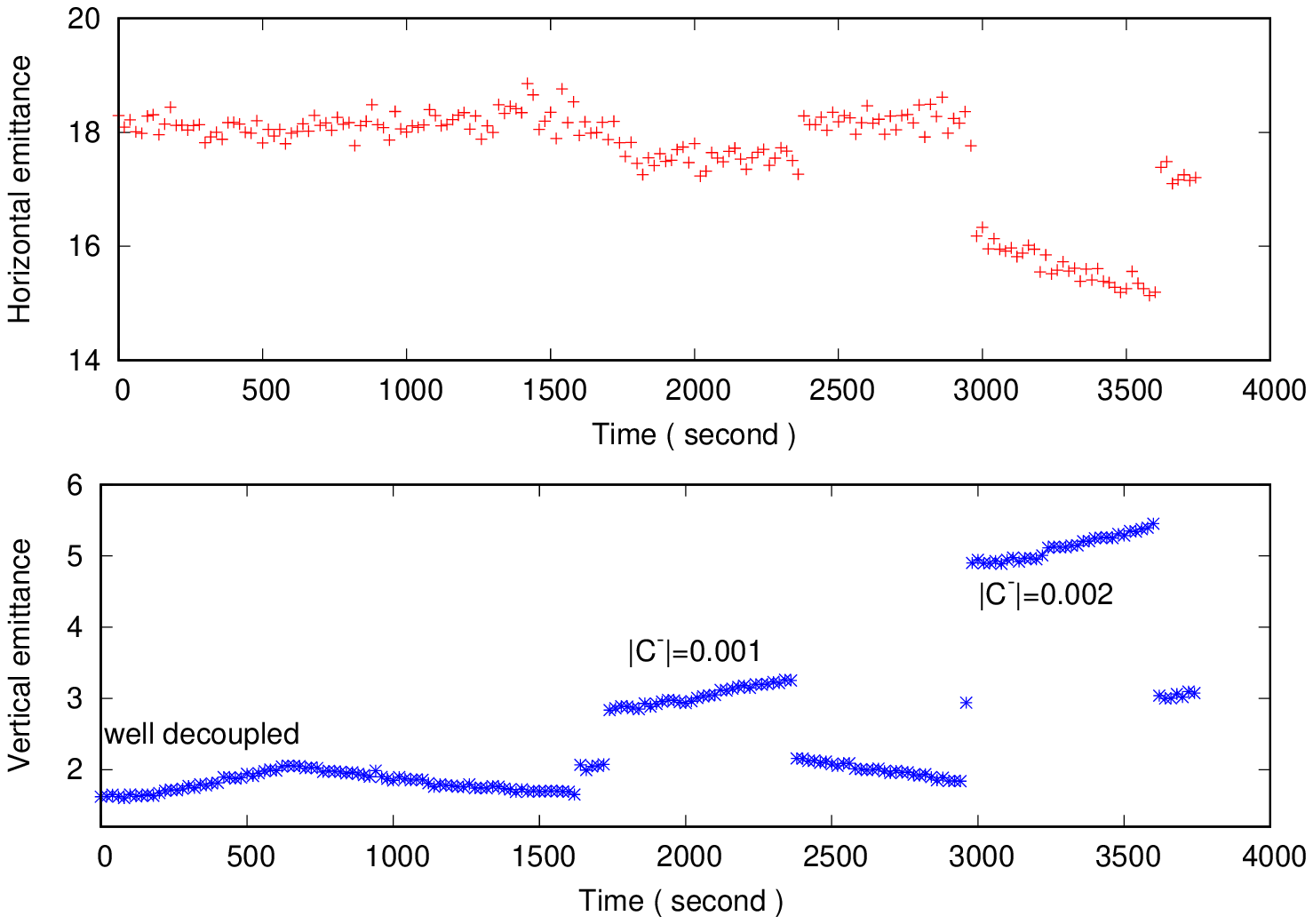}
\caption{\label{fig:5} IPM  Measured  transverse  emittances for the experiment conducted on Nov. 14, 2025.}
\end{figure}
\begin{figure} [hbt]
  \centering
\includegraphics*[width=60mm,angle=-90]{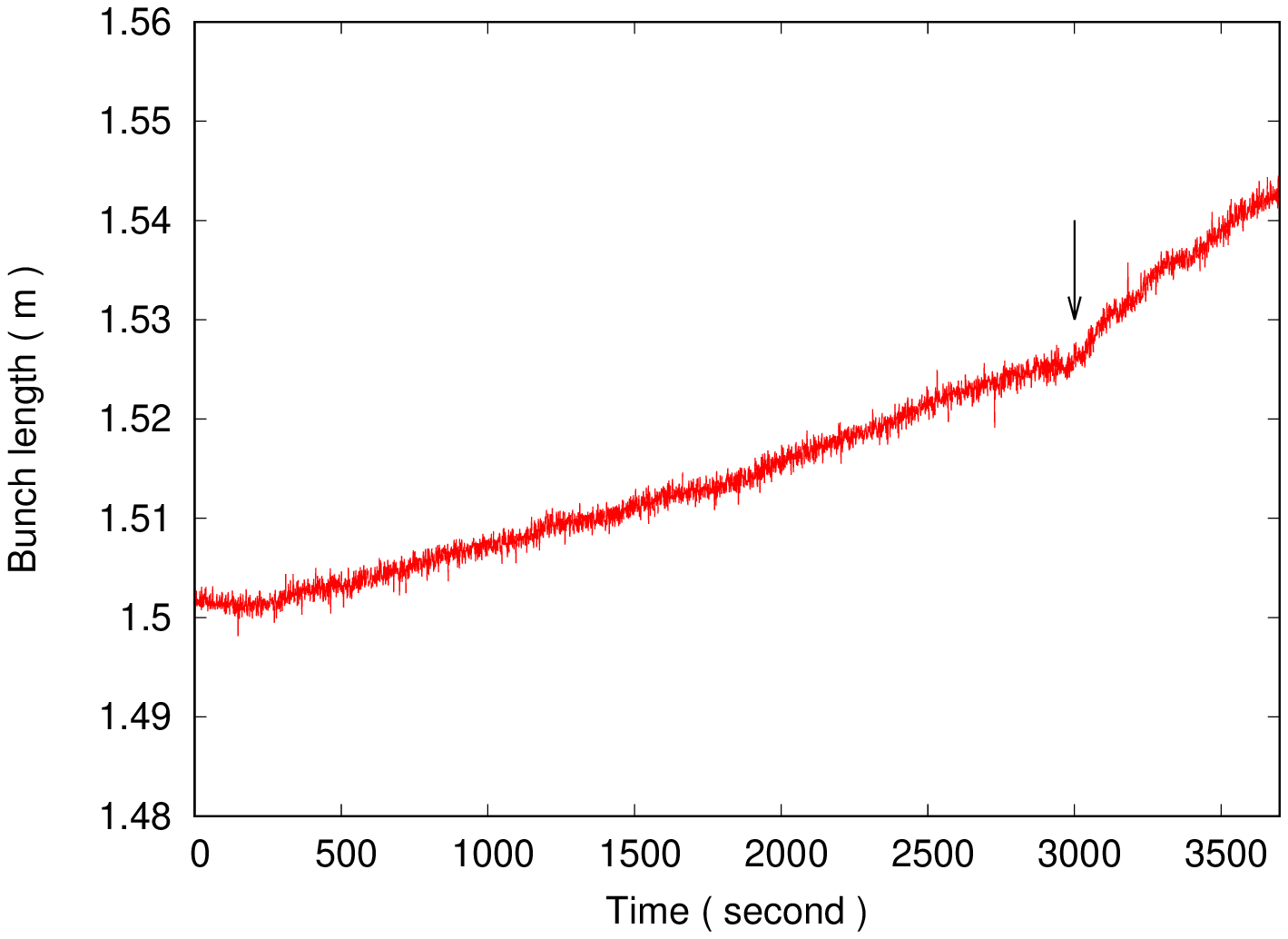}
\caption{\label{fig:6} Measured bunch length for for the experiment conducted on Nov. 14, 2025. }
\end{figure}

Figure~\ref{fig:5} shows the evolution of the quasi-emittances over the full experiment, with the three coupling conditions indicated. The quasi-horizontal emittance remains nearly constant, except in the case with $|C^{-}| = 0.02$, which will be discussed below. In contrast, the quasi-vertical emittance exhibits an immediate increase upon the introduction of coupling, with larger increases observed for stronger coupling. These instantaneous changes arise from coupling effects and have been discussed in detail in Ref.~\cite{PRAB_paper1}. In the following, we focus on the growth of the eigen-emittances and the bunch length in the presence of coupling.

The RMS bunch length is shown in Fig.~\ref{fig:6}. Over approximately one hour, the bunch length increases by only about 3\%, confirming that longitudinal growth remains weak at this energy. A modest increase in the growth rate is observed around 3000~s, coinciding with the ramp-up of the skew-quadrupole currents for the case with $|C^{-}| = 0.02$.

For completeness, IBS modeling was also performed for the well-decoupled condition. The measured and modeled vertical emittance growth times are 36~min and 35~min, respectively, in excellent agreement. The corresponding growth times for the horizontal emittance and the bunch length are effectively infinite.

\subsection{With $|C^{-1}| = 0.01$}

\begin{figure}
  \centering
  \includegraphics*[width=60mm,angle=-90]{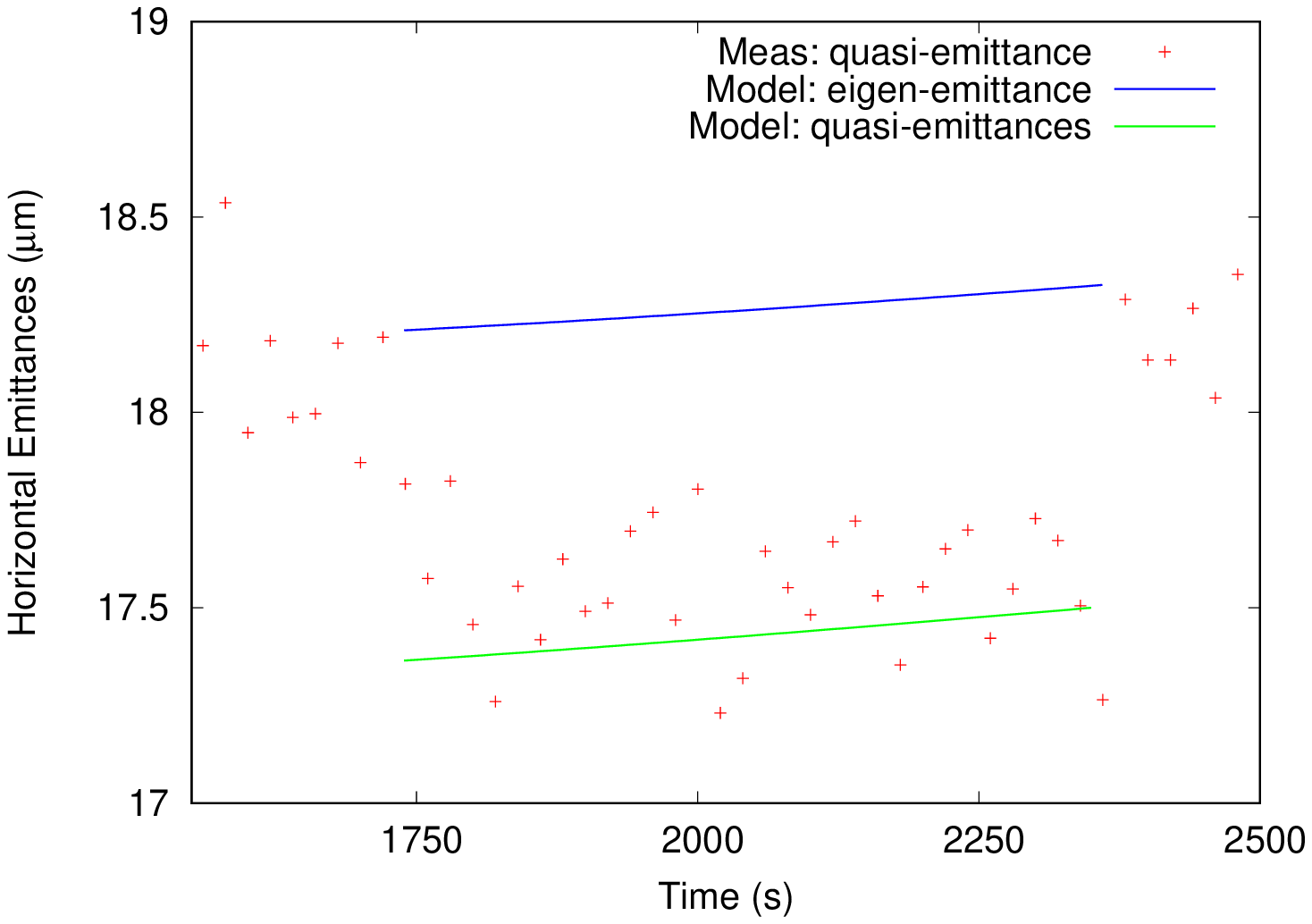}
\caption{\label{fig:7}  Quasi-horizontal emittance and eigenmode I's emittance  with $|C^{-1}| = 0.01$. }
\end{figure}
\begin{figure}
  \centering
  \includegraphics*[width=60mm,angle=-90]{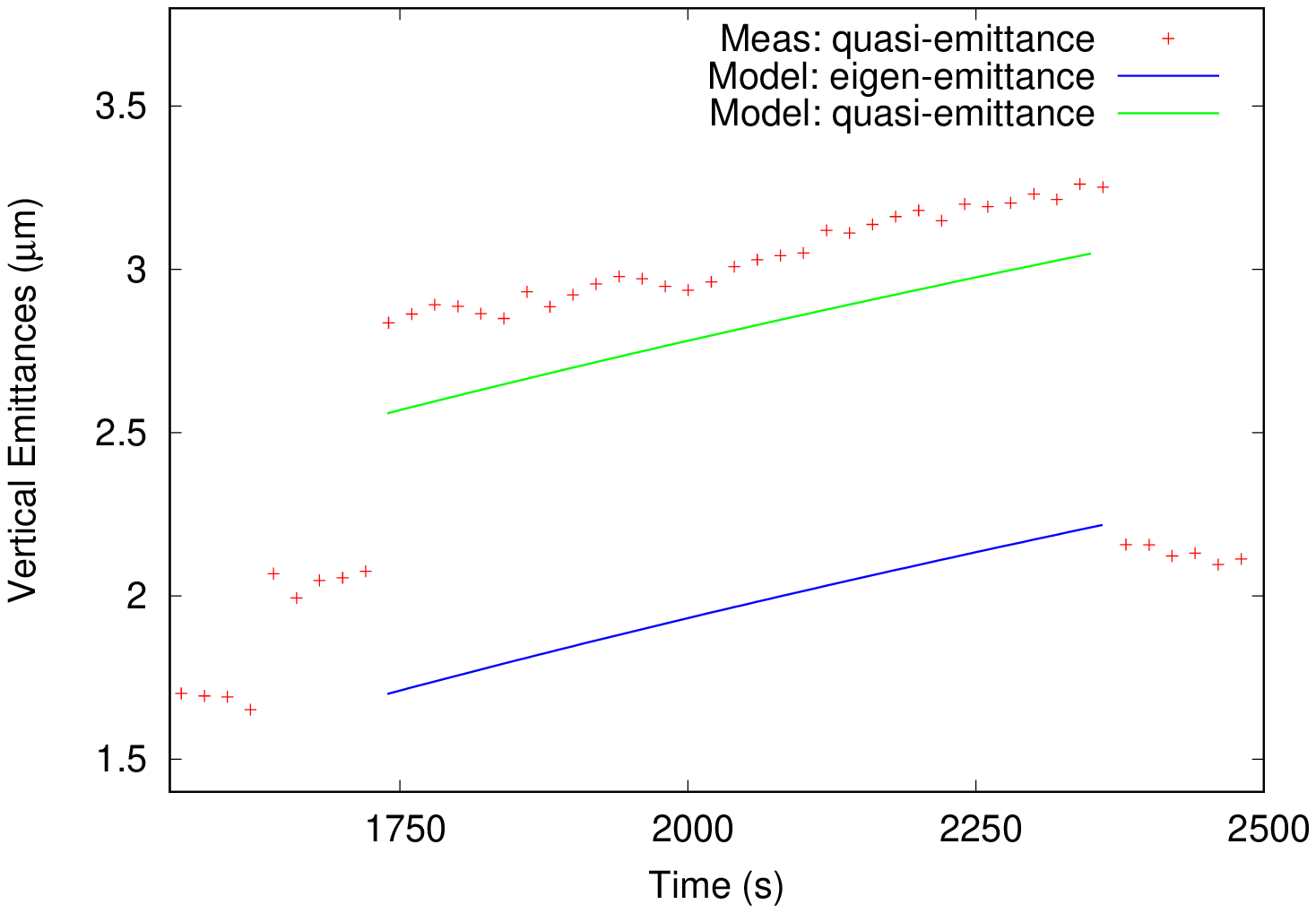}
\caption{\label{fig:8}  Quasi-vertical emittance and eigenmode II's emittance with $|C^{-1}| = 0.01$. }
\end{figure}

In this case, coupling was introduced by changing the skew-quadrupole strengths of the F1 and F3 families, with $\Delta k_{1s} l = 0.1 \times 10^{-3}$. The total change was applied in two equal steps. The amplitude of the coupling coefficient is about an order of magnitude larger than the typical values after applying global decoupling feedback.

Before introducing coupling, the measured fractional tunes were (0.235, 0.215). After coupling was applied, the tunes shifted to (0.236, 0.214). As expected, the betatron tune separation increased with coupling. Using the model lattice, applying the same skew-quadrupole strength changes to the uncoupled lattice yields tunes of (0.2362, 0.2138), in good agreement with the measurements. Unlike IPMs, tune meters measure eigen tunes; in the absence of coupling, these correspond to the horizontal and vertical tunes.

With coupling applied, the bunch-length change remained negligible. Figures~\ref{fig:7} and \ref{fig:8} show the measured quasi-horizontal and quasi-vertical emittances (red dots). The eigen-emittances immediately before coupling was introduced were (18.07~$\mu$m, 1.65~$\mu$m), and those measured immediately after coupling was removed were (18.29~$\mu$m, 2.16~$\mu$m). The mode-I eigen-emittance remained nearly constant over the $\sim$10-minute observation period. The mode-II eigen-emittance increased by approximately 31\%, corresponding to an average growth time of 38~min.

The initial bunch length and momentum spread for IBS modeling are 1.51~m and $1.33 \times 10^{-3}$, respectively. The ion population is $0.696\times10^{9}$ per bunch. To model IBS-induced emittance and bunch-length evolution, we use the IBS formulas of Lebedev–Nagaitsev. In the simulation code, the IBS growth rates, eigen-emittances, bunch length, and momentum spread are updated every 1~s.

For comparison with measurements, the modeled eigen-emittances are also shown as blue curves in Figures~\ref{fig:7} and \ref{fig:8}. The final eigen-emittances from the modeling are (18.17~$\mu$m, 2.18~$\mu$m), in good agreement with IPM measurements immediately after coupling was removed. From the modeling, the mode-II eigen-emittance increases by about 32\% over 10 minutes, corresponding to an average growth time of 37~min.

Using the modeled eigen-emittances $\epsilon_{I,II}$ during the coupled period, the beam sizes at the IPMs and the corresponding quasi-emittances can be reconstructed; these are also shown as green curves in Figures~\ref{fig:7} and \ref{fig:8}. The relative differences between the measured and modeled quasi emittances are  about 1.5\%  and 10\% in  the horizontal and vertical planes. 

For comparison, we also modeled eigen-emittance growth assuming no coupling in the ring. In that case, the mode-II eigen-emittance would increase to 2.21~$\mu$m, corresponding to an increase of about 34\% over 10 minutes. The growth time for eigenmode II without coupling is 35~min. Compared to 37~min with betatron coupling $|C^{-}| = 0.01$, the eigen-emittance growth of mode II is slightly slower. The most significant effect of the coupling is the increase of the quasi-vertical emittance by about 71\%. The ratio of quasi-emittances changes from about 11:1 to 6:1.

\subsection{With $|C^{-1}| = 0.02$}

For this case, stronger transverse coupling was introduced in the ring. The resulting coupling coefficient was approximately 20 times larger than the typical residual coupling after applying decoupling feedback. The skew quadrupole strength changes in the F1 and F3 families were $\Delta k_{1s}l = 0.2 \times 10^{-3}$, also applied in two equal steps.

Prior to introducing coupling, the measured betatron tunes were (0.235, 0.215). After the coupling was applied, the tunes shifted to (0.238, 0.212). The lattice model predicts tunes of (0.2391, 0.2108) for the same skew quadrupole settings, in good agreement with the measurements.

\begin{figure}
  \centering
  \includegraphics*[width=60mm,angle=-90]{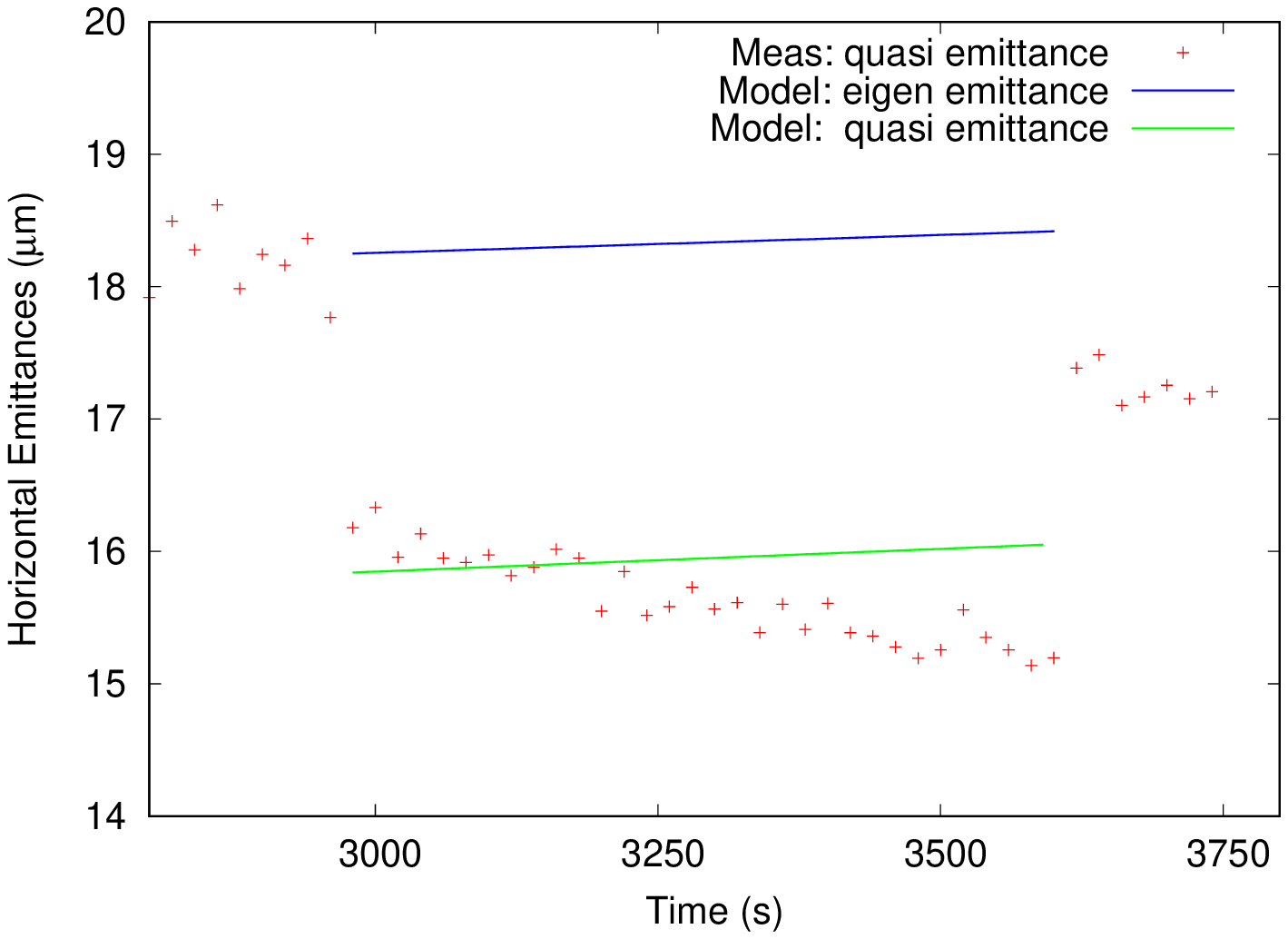}
\caption{\label{fig:9}  Quasi-horizontal emittances and eigenmode I's emittance with $|C^{-1}| = 0.02$. }
\end{figure}

\begin{figure}
  \centering
  \includegraphics*[width=60mm,angle=-90]{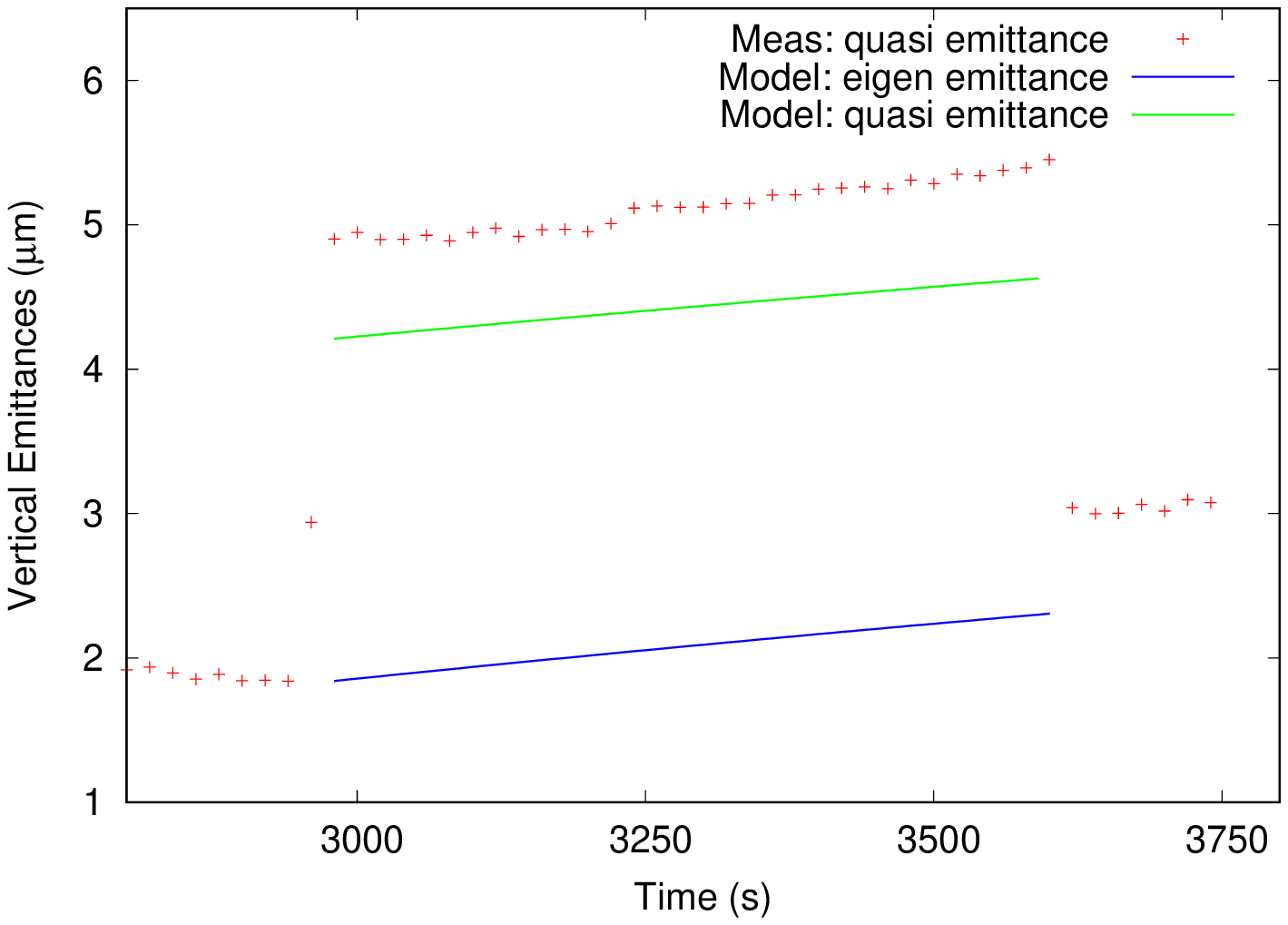}
\caption{\label{fig:10}  Quasi-vertical emittances and eigenmode II's emittance with $|C^{-1}| = 0.02$. }
\end{figure}

Figures~\ref{fig:9} and \ref{fig:10} show the measured quasi-horizontal and quasi-vertical emittances in red. The eigen-emittances immediately before coupling was introduced were (18.25 $\mu$m, 1.84 $\mu$m), while those immediately after coupling was removed were (17.38 $\mu$m, 3.04 $\mu$m). The eigen-emittance of mode I ( mostly associated to horizontal emittance) decreased by about 5\%, whereas that of mode II (mostly associated to vertical emittance) increased by about 65\%.

The initial bunch length and momentum spread for IBS modeling are 1.52~m and $1.33 \times 10^{-3}$, respectively. The ion population is $0.686\times10^{9}$ per bunch.  Using these initial beam parameters, we modeled the evolution of eigen emittances using the IBS formulas by Lebedev and Nagaitsev. In the simulation, the growth rates, eigen-emittances, bunch length, and momentum spread were updated at 1 s intervals. The resulting evolution of the eigen-emittances is shown by the blue curves in Figs.~\ref{fig:9} and \ref{fig:10}.

From the model, over about 10 minutes with coupling, the the eigen-emittances of mode I and mode II would be (18.42 $\mu$m, 2.31 $\mu$m), with relative changes about 0.9\% and 25\%, respectively. The eigen-emittances do not agree with the emittance measurements right after coupling was removed.  Also, significant discrepancies are noticed between the measured and modeled quasi-emittances during the period with coupling.

\begin{figure} [hbt]
  \centering
  \includegraphics*[width=55mm,angle=-90]{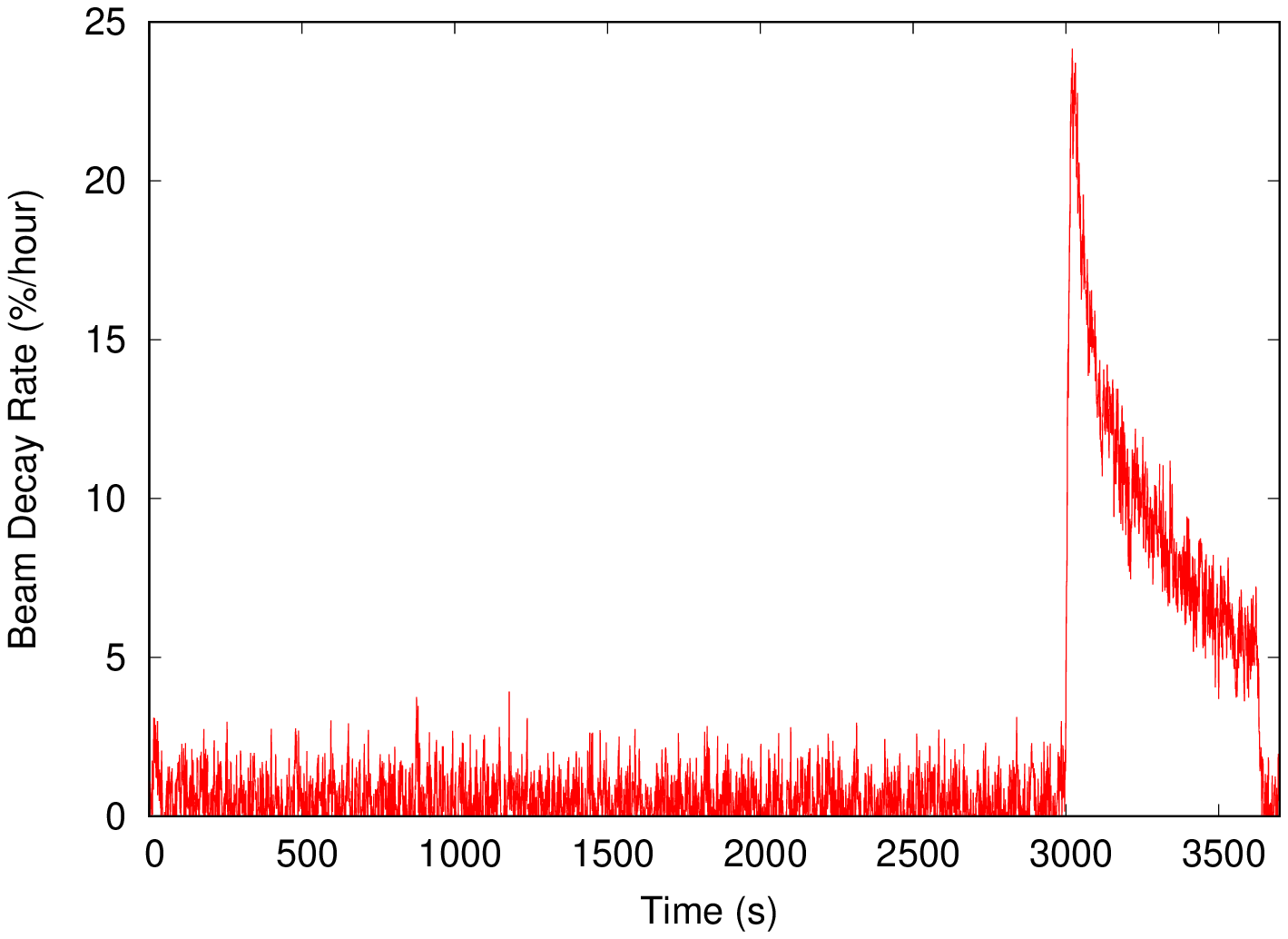}
\caption{\label{fig:11}  Beam decay rate in the whole beam experiment.}
\end{figure}

Figures~\ref{fig:11} show the recorded beam loss rate throughout the experiment. The loss rate remained below 2\%/hour until the skew-quadrupole currents were ramped up for the case with $|C^{-}| = 0.02$, around 3000~s in the plot. At that point, the loss rate peaked at approximately 25\%/hour and then gradually decreased. After the coupling was removed, the beam loss rate returned to its nominal level.

We attribute the observed reduction in the IPM-measured quasi-horizontal emittance to the loss of large-amplitude particles in the horizontal plane. One possible explanation is that strong coupling with $|C^{-}| = 0.02$ pushed the eigentune of mode I close to the fourth-order resonance at 0.25. A similar effect has been consistently observed during the early phase of physics stores in proton–proton collisions at RHIC, where large-amplitude particles are lost due to reduced momentum aperture in the presence of beam–beam interactions. There we also observed IPM measured transverse emittance reduction in the first hour of stores~\cite{RHIC-BB}.

\begin{figure}[hbt]
  \centering
  \includegraphics*[width=60mm,angle=-90]{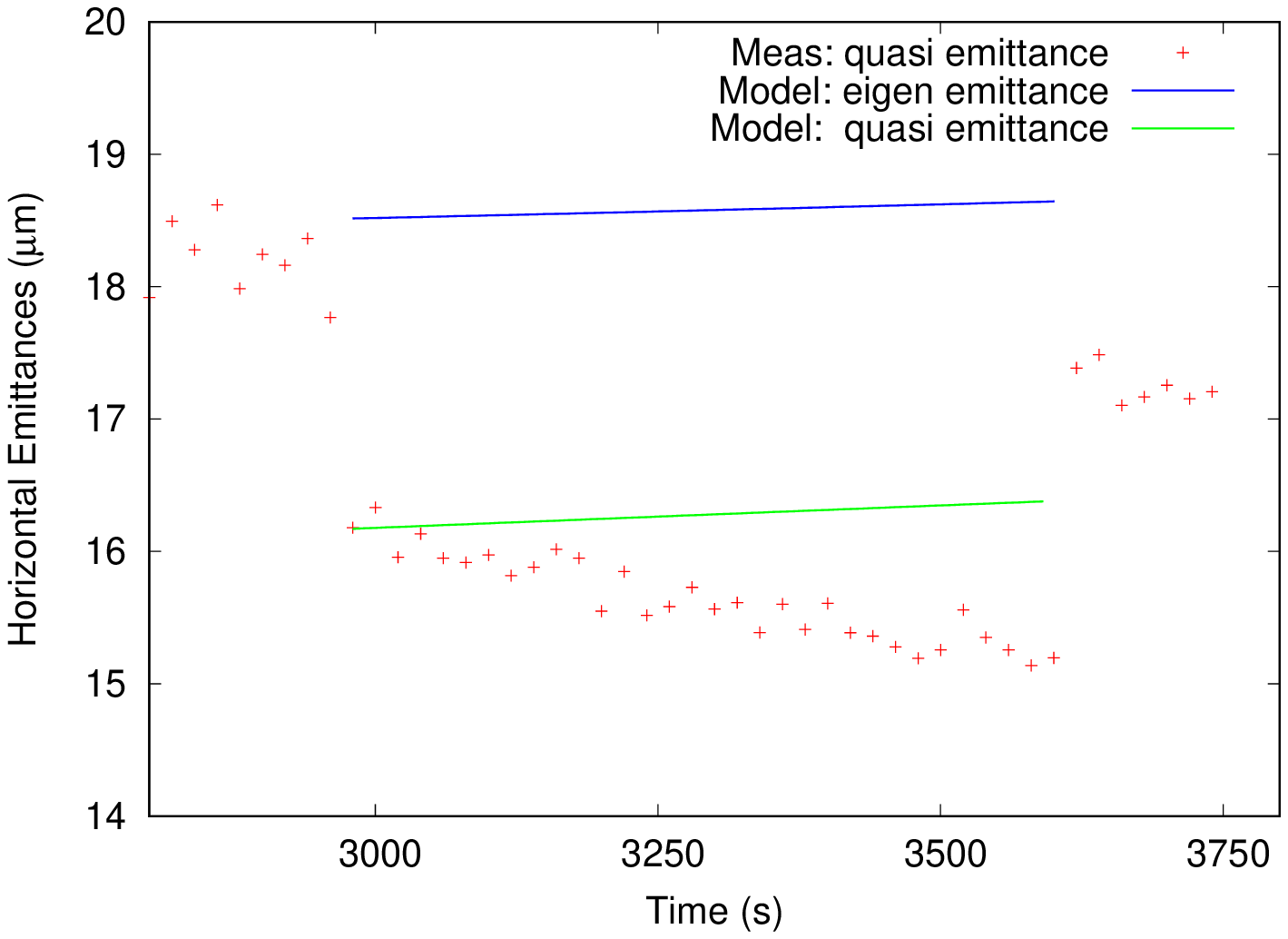}
\caption{\label{fig:12}  Comparison of quasi-horizontal emittances from measurement and eigenmode I'S emittance from IBS modeling. }
\end{figure}

\begin{figure}[hbt]
  \centering
  \includegraphics*[width=60mm,angle=-90]{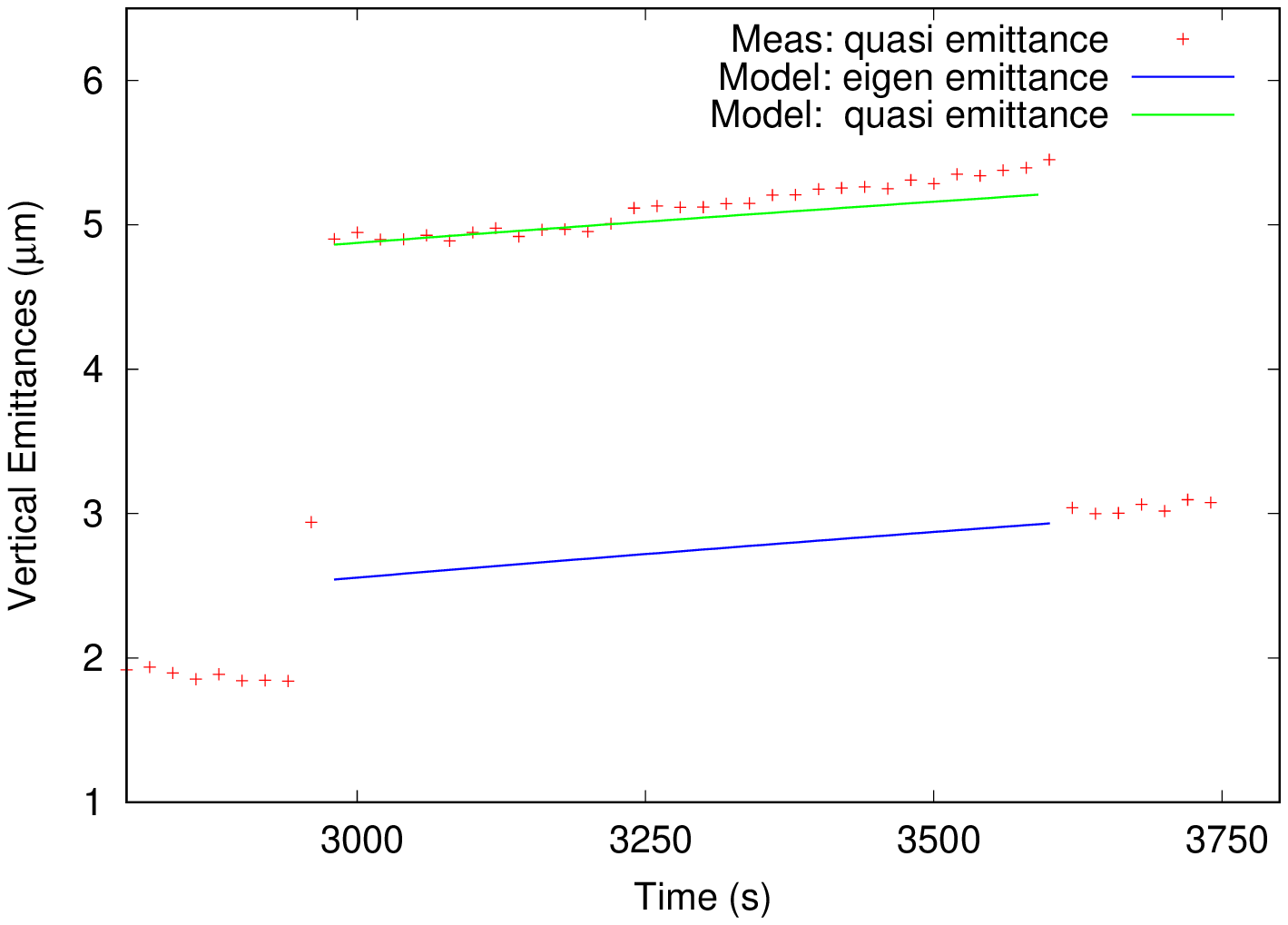}
\caption{\label{fig:13}  Comparison of quasi-horizontal emittances from measurement and eigenmode I'S emittance from IBS modeling. }
\end{figure}

Comparing the eigenmode-II emittances from measurement and modeling after coupling was removed, we infer that additional emittance increase occurred during the ramp-up of the skew-quadrupole currents. Relative to the case with $|C^{-}| = 0.01$, the ramp rate was approximately twice as fast, which may have induced non-adiabatic eigen-emittance growth.

In addition, small discontinuities in the measured quasi-emittances are observed around 3200~s in Figs.~\ref{fig:9} and \ref{fig:10}, which may have contributed to additional increases in the eigen-emittances. At 3200~s, an unplanned automatic store orbit correction was triggered.

To better represent the initial beam parameters for IBS modeling with coupling, we adjusted the initial eigen-emittances to match the IPM-measured quasi-emittances of (16.20~$\mu$m, 4.85~$\mu$m) immediately after the introduction of coupling. An optimization procedure was used to obtain the corresponding eigen-emittances of (18.51~$\mu$m, 2.54~$\mu$m). Compared to the measured values prior to coupling (18.25~$\mu$m, 1.87~$\mu$m), this implies an increase of approximately 36\% in the eigenmode-II emittance. These results suggest that slower ramping of the skew-quadrupole currents is required to preserve the eigen-emittances.

Using these adjusted initial eigen-emittances (18.51~$\mu$m, 2.54~$\mu$m), we re-modeled the evolution of the eigen-emittances and bunch length for the coupling case $|C^{-}| = 0.02$, as shown in Figs.~\ref{fig:12} and \ref{fig:13}. With the updated initial conditions, the IBS model reproduces the IPM-measured eigenmode-II emittance immediately after coupling removal, with a relative error of 4\%. The modeled quasi-vertical emittance also agrees well with the measurements, although a small increase is observed around 3200~s due to the automatic store orbit correction. As noted above, the decrease in eigenmode-I emittance is attributed to particle loss in the horizontal plane and is therefore not captured by the model.

For comparison, with initial eigen-emittances of (18.25~$\mu$m, 1.84~$\mu$m), the modeled growth time for eigenmode II is about 46 minutes. Compared to about 37 minutes in the case above with $|C^{-}| = 0.01$, the emittance growth of eigenmode II is slower. Increasing the coupling in the ring leads to slower growth of eigenmode II. In our experiments, the initial eigen-emittance of mode II was about 10 times smaller than that of mode I. Again, the most significant effect of the coupling is an increase in the quasi-vertical emittance. The ratio of quasi-emittances changes from about 10:1 to 3:1 when $|C^{-}| = 0.02$.

\section{Implications for the EIC}

In this section, we first evaluate the IBS growth times for the EIC/HSR at various beam energies. We then study the impact of the transverse emittance ratio and betatron coupling on the IBS growth time. The IBS growth times are calculated using the latest proton design lattices, and the Lebedev–Nagaitsev formalism is used.

\subsection{IBS growth times for the HSR}

The EIC/HSR will reuse the arcs of the yellow ring of RHIC and connect them with new straight sections. The hadron injection accelerators for RHIC will be retained for the EIC/HSR. The typical 95\% normalized transverse emittance for the RHIC injection accelerators is about 15~$\mu$m.

For proton–electron collisions in the EIC, the ratio of transverse beam sizes at the IP is about 10:1 to achieve high peak luminosity. The proton beam will first be cooled at the injection energy of about 23.8~GeV using conventional electron cooling~\cite{LEREC}. The transition energy for the injection lattice is 23.5~GeV. After a large emittance ratio is achieved, the flat proton beam will be accelerated to various collision energies.

Table~\ref{tab:3} lists the design parameters at the injection energy and at two collision energies, 100 and 275~GeV. For consistency throughout this paper, we use 95\% normalized transverse emittances in this table. The parameters at injection correspond to the cooled proton beam. The IBS growth times $\tau_{x,y,z}$ correspond to the transverse and longitudinal emittances, respectively.

The proton bunch intensity at injection energy is $28 \times 10^{10}$. This bunch will be split into four bunches at collision energies. The IBS growth rate is proportional to the bunch intensity and also strongly depends on the particle energy. As shown in Table~\ref{tab:3}, the IBS growth time for vertical emittance is only 18 minutes for the cooled beam. It is crucial to accelerate the beam immediately after the design emittances are reached.

From Table~\ref{tab:3}, with the increase in the proton energy to collision energies, the growth time for the vertical emittance gradually increases and becomes to infinite at 275~GeV. The reason is that there is no vertical dispersion by design  for the EIC/HSR. The vertical emittance increase with IBS will come from the horizontal emittance growth and betatron coupling.

\begin{table} 
	\vspace*{-.8\baselineskip}
	\centering 
	\caption{\label{tab:3} Design parameters for proton beams and calculated IBS growth times}
	\begin{tabular}{lcccc} 
		\hline \hline
		\bf quantities   &    unit      &     \bf 23.8~GeV         &  \bf  100~GeV      &   \bf  275~GeV   \\  \hline
                 $\gamma$        &     -        &      25.4                &  106.6              &     293.1         \\
                 Bunch intensity &    $10^{10}$ &       28                  &   6.9              &     6.9            \\
                 Hori. emittance &   $\mu$m     &      18.7                 &   19.2              &     19.9              \\
                 Vert. emittance &   $\mu$m     &      1.7                 &   1.72              &     1.76              \\
		Bunch length     &    m         &      1                   &   0.07              &     0.06             \\
		Momentum spread  &    $10^{-4}$  &      6                   &   9.7               &     6.8                  \\ 
                IBS $\tau_x$     &    hour         &   -6.8                 &   1.7               &     2.1   \\
                IBS $\tau_y$     &    hour         &   0.3                  &   4.1               &     +INF  \\
                IBS $\tau_z$     &    hour         &   0.67                 &   3.0               &     3.9\\
                \hline  \hline
	\end{tabular} 
\end{table}

\subsection{IBS vs transverse emittance ratio}

For electron cooling at the injection energy, it is more efficient to cool both transverse planes simultaneously rather than cooling only the vertical plane. It takes about 30 minutes to cool the transverse emittances to (0.5~$\mu$m, 0.3~$\mu$m). Table~\ref{tab:4} compares the IBS growth times for different transverse emittance ratios $\kappa$. In this calculation, the vertical emittance is kept constant. The transverse emittance ratio is changed by adjusting horizontal emittance.

From Table~\ref{tab:4}, a larger initial horizontal emittance or a larger transverse emittance ratio at injection leads to slightly faster growth times for the vertical emittance and the bunch length. Based on simulation, the horizontal emittance shrink is seen when the emittance ratio is larger than 6:1.  The emittance ratio 1.7:1 or 5:3 was proposed by the cooling design team. If we  cool the proton beams in both transverse planes, the horizontal emittance need to blow up at the collision energy to match the design emittance ratio for collision.

\begin{table} 
	\vspace*{-.8\baselineskip}
	\centering 
	\caption{\label{tab:4} IBS growth times with different transverse emittance ratio at proton's injection energy. For this calculation, the vertical emittance is kept constant. }
	\begin{tabular}{lccc} 
		\hline \hline
	         $\kappa=\epsilon_x / \epsilon_y$         &    IBS $\tau_x$     &    IBS $\tau_y$     &   IBS $\tau_z$    \\
                                                         &     (  hour )         &   (  hour )      &  (  hour  )    \\   \hline
                11       &                  -6.8         &          0.3          &        0.67         \\
                8        &                  -5.5         &          0.26         &        0.60         \\
                6        &                  -5.2         &          0.24          &       0.56         \\
                4        &                  -14.0         &           0.22         &        0.50        \\
                1.7      &                  0.4         &           0.24          &       0.41         \\
                \hline \hline
	\end{tabular} 
\end{table}

At the collision energy, a larger transverse emittance ratio leads to higher peak luminosity. Table~\ref{tab:5} compares the calculated IBS growth times at 275~GeV for transverse emittance ratios of 11:1, 8:1, 6:1, and 4:1. In this calculation, the horizontal emittance is kept constant. The transverse emittance ratio is changed by adjusting vertical emittance.

The results show that a smaller transverse emittance ratio increases the IBS growth times of the horizontal and longitudinal emittances, while the vertical IBS growth time remains effectively infinite. With a high transverse emittance ratio, say 11:1, beam-beam interaction simulations indicate that the proton beam will have  a larger vertical emittance growth rate and a smaller dynamic aperture. Therefore, it is necessary to balance the transverse emittance ratio ( peak luminosity) and beam lifetime to maximize the integrated luminosity.

\begin{table} 
	\vspace*{-.8\baselineskip}
	\centering 
	\caption{\label{tab:5} IBS growth times with different transverse emittance ratio at proton's 275 GeV. For this calculation, the horizontal emittance is kept constant.}
	\begin{tabular}{lccc} 
		\hline \hline
	         $\kappa=\epsilon_x / \epsilon_y$         &    IBS $\tau_x$     &    IBS $\tau_y$     &   IBS $\tau_z$    \\
                                                         &     (  hour )         &   (  hour )      &  (  hour  )    \\   \hline
                11       &                  2.1         &           +INF         &         3.9         \\
                8        &                  2.5         &           +INF         &         4.7         \\
                6        &                  3.0         &           +INF         &         5.6         \\
                4        &                  3.8         &           +INF         &         7.1          \\                
                \hline \hline
	\end{tabular} 
\end{table}

\subsection{IBS vs betatron coupling}

The sources of betatron coupling in the HSR include the detector solenoid, quadrupole roll angles, and vertical closed orbits at arc sextupoles. The full length of the EIC detector solenoid in IR6 is 4~m, with a design longitudinal magnetic field of about 2~T. The strength of the detector solenoid scales inversely with particle energy. Its contribution to the global coupling coefficient is 0.053 at 23.8~GeV and 0.0053 at 275~GeV. As discussed in Ref.~\cite{PRAB_paper1}, the amplitude of the coupling coefficient should be less than 0.002 to achieve an 11:1 transverse emittance ratio.

For a random distribution of quadrupole roll angles along the ring, the amplitude of the coupling coefficient can be estimated as
\begin{equation}
|C^{-}|^2 = \frac{1}{\pi^2} \left( \sum \beta_x \beta_y (k_1 l)^2 \right) \langle \theta^2_{\text{roll}} \rangle.
\end{equation}
Here $\langle \theta_{\text{roll}} \rangle$ is the RMS of the quadrupole roll errors. Its contribution is not sensitive to particle energy. For example, using the latest 275~GeV HSR store lattice and assuming $\langle \theta_{\text{roll}} \rangle = 100$,\textmu rad, the resulting $|C^{-}|$ is 0.0054.

As discussed in Ref.~\cite{PRAB_paper1}, the amplitude of the coupling coefficient should be about one order of magnitude smaller than the transverse tune split. For example, for the HSR design tunes (0.228, 0.210) at collision energy, $|C^{-}|$ should be less than 0.0018. We will employ the same RHIC decoupling feedback for the EIC/HSR. As mentioned earlier, the typical coupling amplitude in RHIC is about 0.001 after applying decoupling feedback. During cooling at injection and acceleration, the tune split can be increased, provided that the beam lifetime remains acceptable.

\begin{figure} 
  \centering
  \includegraphics*[width=60mm, angle=-90]{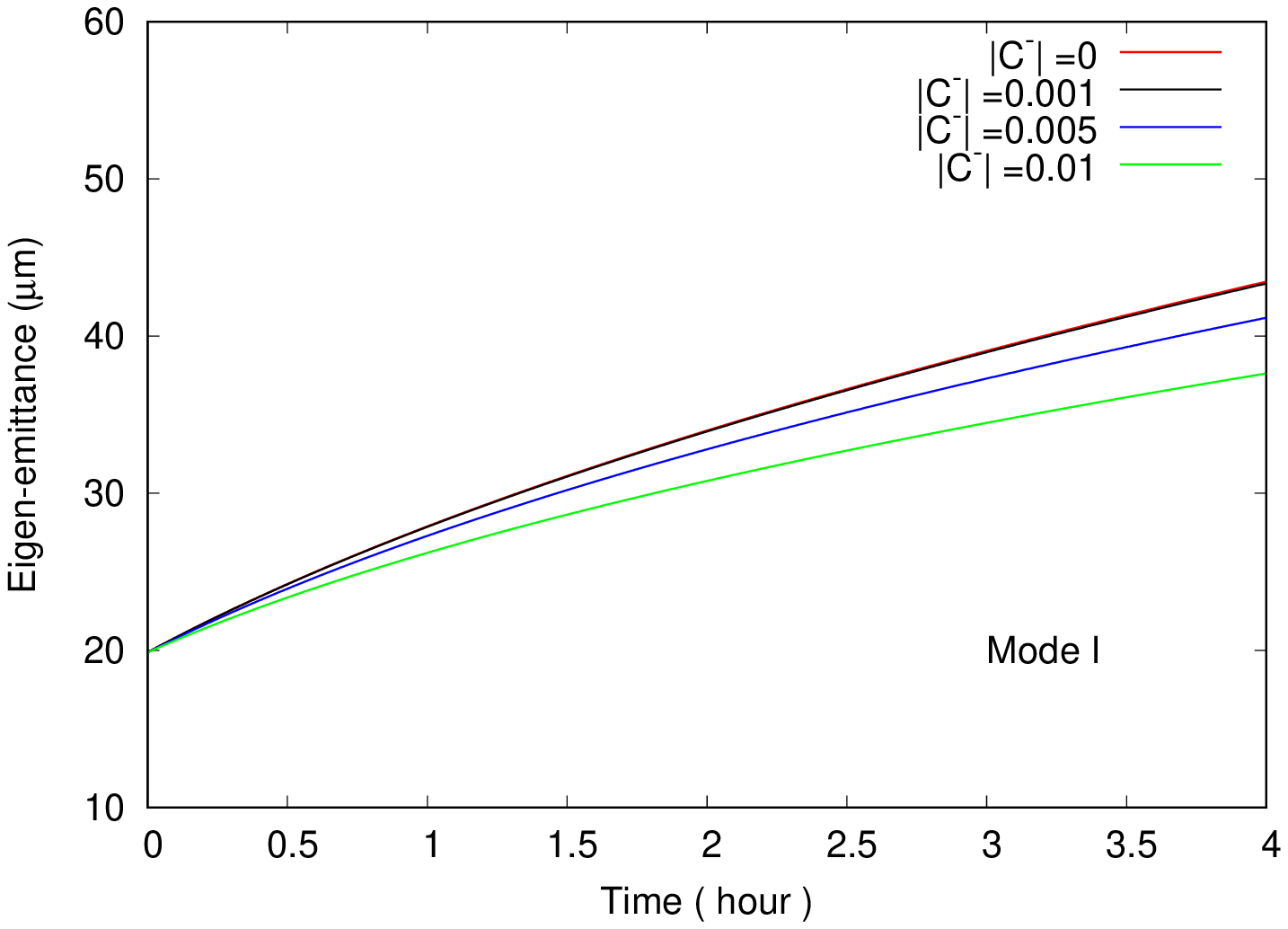}
\caption{\label{fig:15}   Evolution of eigen-emittance of mode I with different coupling for proton beam at  275~GeV.}
\end{figure}

\begin{figure} 
  \centering
  \includegraphics*[width=60mm,angle=-90]{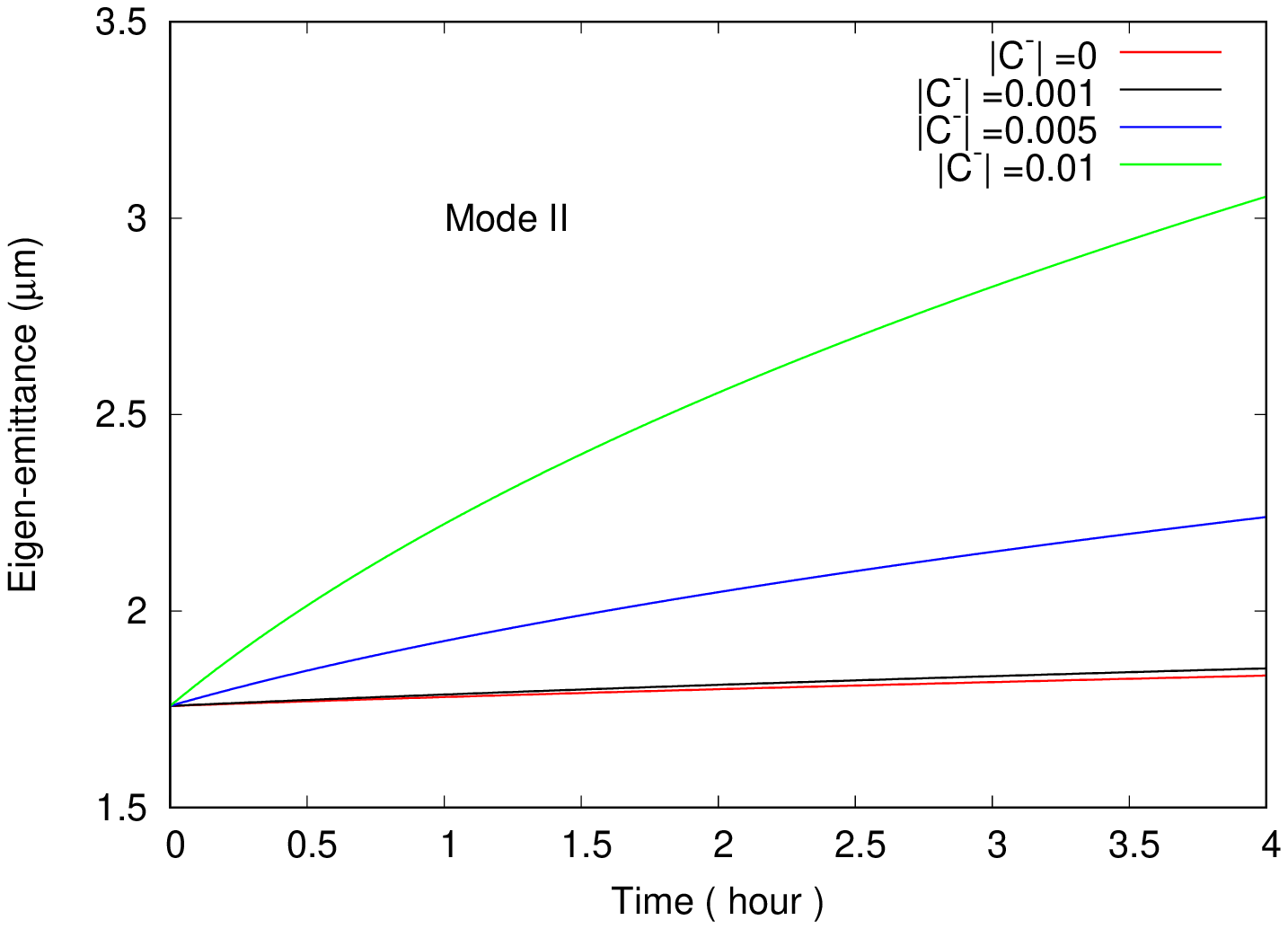}
\caption{\label{fig:16}   Evolution of eigen-emittance of mode II with different coupling for proton beam at 275~GeV. }
\end{figure}

Here we simulate the evolution of eigen-emittances in the presence of residual weak betatron coupling. Using the 275~GeV HSR store lattice, coupling is introduced through quadrupole rolls. The transverse tunes without coupling are set to (0.228, 0.210). In the simulation, the IBS growth rates and eigen-emittances are updated every 10 seconds. Figures~\ref{fig:15} and~\ref{fig:16} show the eigen-emittances of mode I and mode II under four different coupling conditions: no coupling, $|C^{-}| = 0.001$, $|C^{-}| = 0.005$, and $|C^{-}| = 0.01$. The simulation is performed over a real beam time of 4 hours.

From the plots, as the betatron coupling in the ring increases, the IBS growth rate for eigenmode II increases, while the IBS growth rate for eigenmode I decreases. Comparing the cases without and with coupling, the relative change in the eigen-emittance of mode I is much smaller than that of mode II. There is about a 5\% increase in the eigen-emittance of mode II over 4 hours  if there is not coupling in the ring. There is no significant difference in the evolution of eigen-emittances between the cases without coupling and $|C^{-}|=0.001$. For $|C^{-}|=0.005$, the eigen-emittance of mode II increases by about 32\% over the course of the simulation. The enlarged eigen-emittance of mode II increases the vertical beam size at IP and therefore reduces luminosity.

In our beam experiment at 31 GeV/nucleon, we observed a slower emittance growth rate for eigenmode II after introducing coupling in the ring. The reason is that, at low energy, the dominant IBS growth occurs in the vertical plane due to the large transverse emittance ratio, while horizontal IBS growth is negligible. When coupling is introduced, the IBS-driven emittance growth of eigenmode II is partially transferred to eigenmode I.  In contrast, for a 275 GeV proton beam, the dominant IBS-induced emittance growth occurs in the horizontal plane, with negligible vertical growth. After coupling is introduced, the IBS-driven emittance growth of eigenmode I is partially transferred to eigenmode II. The main impact of coupling on IBS is to mix or redistribute the emittances and their growth.

As discussed in Ref.~\cite{PRAB_paper1}, emittance growth from IBS in the presence of coupling cannot be simply interpreted as a process in which diffusion occurs only in one transverse plane (horizontal or vertical), while the emittance growth in the other plane is entirely transferred from the IBS-dominant plane through coupling. In reality, coupling modifies the transverse particle distribution, which in turn affects the IBS growth rates.

\section{Summary}

In this article, we benchmark IBS-induced emittance growth for beams with large transverse emittance ratios in RHIC, both without and with betatron coupling, using the Lebedev–Nagaitsev formalism. The simulations reproduce the observations for uncoupled cases at 31~GeV/nucleon and 100~GeV/nucleon, as well as for the weakly coupled case with $|C^{-}| = 0.01$.

For $|C^{-}| = 0.02$, strong coupling pushes the eigenmode~I tune close to a fourth-order resonance, resulting in particle loss and a reduction of the eigenmode~I emittance. The eigenmode~II emittance is likely not an adiabatic invariant during the skew quadrupole ramp. By adjusting the initial eigen-emittances in the IBS model to match the quasi-emittances, the observed evolution of the eigenmode~II emittance can be reproduced.

IBS growth times are also evaluated for proton beams in the EIC/HSR over a range of energies. To achieve and maintain large emittance ratios for high-luminosity collisions, betatron coupling must be minimized. For the EIC/HSR, RHIC-like decoupling feedback will be employed to suppress the coupling amplitude to below 0.002.

\section{Acknowledgment}

This work was supported by the Office of Science, U.S. Department of Energy, under contracts DE-SC0012704 and DE-AC05-06OR23177. We express our gratitude to the RHIC Accelerator Physics EXperiment (APEX) program for providing valuable beam time for our experiments, and we thank the RHIC Main Control Room operation crew for their support during these experiments.

\bibliography{bibliography} 

\end{document}